\documentclass[preprint,showpacs,preprintnumbers,amsmath,amssymb]{revtex4}

\usepackage{graphicx}
\usepackage[utf8x]{inputenc}
\usepackage{color}
\usepackage{subfig}
\usepackage{multirow}

\usepackage{epsfig}

\begin{document}

\title{Supersymmetric $U(1)_{Y^{\prime}}\otimes U(1)_{B-L}$ extension of the standard  model}

\author{J. C. Montero}
\email{montero@ift.unesp.br}
\affiliation{
	Instituto  de F\'\i sica Te\'orica--Universidade Estadual Paulista \\
	R. Dr. Bento Teobaldo Ferraz 271, Barra Funda\\ S\~ao Paulo - SP, 01140-070,
	Brazil
}
\author{V. Pleitez}%
\email{vicente@ift.unesp.br}
\affiliation{
	Instituto  de F\'\i sica Te\'orica--Universidade Estadual Paulista \\
	R. Dr. Bento Teobaldo Ferraz 271, Barra Funda\\ S\~ao Paulo - SP, 01140-070,
	Brazil
}

\author{M. C. Rodriguez}%
\email{marcoscrodriguez@ufrrj.br}
\affiliation{ Grupo de F\'\i sica Te\'orica e Matem\'atica F\'\i sica \\
	Departamento de F\'\i sica,
	Universidade Federal Rural do Rio de Janeiro \\
	BR 465 Km 7, 23890-000, Serop\'edica, RJ, Brazil
}

\author{ B. L. S\'anchez-Vega}%
\email{brucesan@ift.unesp.br}
\affiliation{
	Instituto  de F\'\i sica Te\'orica--Universidade Estadual Paulista \\
	R. Dr. Bento Teobaldo Ferraz 271, Barra Funda\\ S\~ao Paulo - SP, 01140-070,
	Brazil
}

\date{\today}

\begin{abstract}
We build a  supersymmetric version with  $SU(3)_C\otimes SU(2)_L\otimes U(1)_{Y^\prime}\otimes U(1)_{B-L}$ gauge symmetry, where $Y^\prime$ is a new charge and $B$ and $L$ are the usual baryonic and leptonic numbers.
The model has three right-handed neutrinos with identical $B-L$ charges, and can accommodate all fermion masses at the tree level. In particular, the type-I seesaw mechanism is implemented for the generation of the active neutrino masses. We  obtain  the mass spectra of all  sectors and for the scalar one we also give the flat directions allowed by the model.
\end{abstract}

\pacs{12.60.Jy  
12.60.Cn 
}

\maketitle
\section{Introduction}
\label{sec:intro}

Although the Standard Model (SM) is in agreement with the
observed properties of the charged fermions, it is unlikely to be the
ultimate theory. From the experimental point of view, we can mention at least three evidences indicating that there must be physics beyond the SM: i) The nonzero neutrino masses~\cite{Agashe:2014kda}. In the SM the neutrinos are massless to all orders in perturbation theory, and non-perturbative effects are negligible, at least at zero temperature;
ii) The strength of CP violation in the single phase in the Cabibbo-Kobayashi-Maskawa
(CKM) mixing matrix,  is not able to generate the observed baryon asymmetry in the
universe~\cite{Bernreuther:2002uj}; iii) The existence of dark matter (DM)~\cite{Gaskins:2016cha}, in the SM there is no such a candidate to explain this issue.

One of the most popular extensions of the SM is the Minimal Supersymmetric
Standard Model (MSSM). See~\cite{Fayet:2001xk,Rodriguez:2009cd} and references therein.
It is usual in the MSSM framework to  introduce  $R$-parity in order to forbid baryon and lepton number violating interactions
avoiding  proton decay  and processes like
$\mu \to e \gamma$.
When the model satisfies the $R$-parity the lightest supersymmetric particle (LSP) is stable and, therefore, it is
a candidate to be the dark matter in the Universe \cite{drees,tata}.
However, in this model  neutrinos are massless. If we want
to generate masses to them we need to break the $R$-parity generating violation of
the lepton number $L$, but baryon number $B$ is still conserved, so that the proton remains stable and at least one neutrino gets mass at the tree
level \cite{Hall:1983id,Montero:2001ch}.
Moreover, the MSSM contains new CP-violating sources beyond the CKM matrix~\cite{tata}. Therefore, mechanisms for generating baryogenesis through leptogenesis~\cite{Fukugita:1986hr}
are natural in supersymmetric models~\cite{Kajiyama:2009ae}.

It is a well known fact that in the context of the SM (no right-handed neutrinos)  both,
baryon  and lepton  numbers, are conserved automatically. However, both $U(1)_B$ and $U(1)_L$ are
anomalous~\cite{'tHooft:1976up} (but their consequences are well suppressed,
at least at zero temperature) and only the combination $U(1)_{B-L}$ is
anomaly free. When  the $U(1)_{B-L}$ symmetry is gauged, it
becomes anomaly free provided an appropriate number of right-handed neutrinos
is added, for instance, one per family, as in the left-right symmetric models~\cite{Marshak:1979fm}.

Both neutrino masses and the baryon asymmetry problems can be readily solved if
right-handed neutrinos are introduced. Once right-handed neutrinos are introduced the baryon number  minus lepton number,  i.e.,
the $B-L$ quantum number can be related to a local $U(1)$ gauge symmetry. A possibility is to add an extra $U(1)$ factor to the SM gauge symmetry, denoted by $G_{\textrm{SM}}$. The symmetry at high energy would be $G_{B-L}\equiv G_{\textrm{SM}}\otimes U(1)_{B-L}$, and $G_{B-L}\to G_{\textrm{SM}}$ after the  symmetry is broken at the TeV scale~\cite{Emam:2007dy}.

If we consider $U(1)_{B-L}$ as an additional gauge symmetry  we can introduce a term such as $\sim N_{i}N_{j}\phi$ if $L( \phi)=-2$, where $N_i$ denotes the right-handed neutrinos, and a Majorana mass term is induced to the neutrinos if $\langle\phi\rangle\not=0$. If this term is not present, there is a fine tuning since the respective Yukawa couplings are severely constrained, ${\lesssim}10^{-12}$, in order to explain the Dirac neutrino masses, see for instance Ref.~\cite{Allahverdi:2007wt}.

Recently it was proposed  a model in which $U(1)_{B-L}$ is not just a new factor added to the SM gauge symmetry but $U(1)_Y$ in $G_{\textrm{SM}}$ is substituted   by $U(1)_{Y^\prime}\otimes U(1)_{B-L}$  and the breaking $U(1)_{Y^\prime}\otimes U(1)_{B-L}\to U(1)_Y$ occurs at the TeV scale~\cite{Montero:2007cd}.
Moreover, the number of right-handed neutrinos and their $B-L$ (or $Y^\prime$) quantum numbers are free parameters,  but the  cancellation of the cubic and the linear anomalies implies that at least  three right-handed neutrinos must be added to the matter representation content of the SM. Explicit solutions for the $B-L$ (or $Y^\prime$) parameters show that at least two types of model arise: i) the model with three right-handed neutrinos having  $B-L=-1$; ii) the model with two right-handed neutrinos having $B-L=-4$ and the third one having $B-L=5$. In models of the second type it is possible to generate Dirac masses or the inverse seesaw mechanism for the active neutrinos~\cite{Ma:2014qra}. Phenomenology of both sort of models were consider in Refs. \cite{Fortes:2010mz,Coutinho:2011xb,Montero:2011jk,Bruce:2014DM,Bruce:2015DM}.
Here we will consider only the supersymmetric extension of the model in which the right-handed neutrinos that are sterile with respect to the $G_{\textrm{SM}}$ interactions have the same $B-L$ as the active neutrinos.

The outline of this paper is as follows. In
Sec.~\ref{sec:particles} we present the particle content of the model with the respective quantum numbers, and the Lagrangian  in Sec.~\ref{sec:lagrangian}. There  we define the $R$-parity, and in Subsec.~\ref{subsec:susyterms} we give the supersymmetric Lagrangian, while the soft term is shown in Subsec.~\ref{subsec:soft}.  In Sec.~\ref{sec:masses} we obtain the gauge boson and fermion mass spectra of the model. The scalar potential is the theme of Sec.~\ref{sec:potential} and the mass spectra in the scalar sectors are given in Sec.~\ref{sec:higgs}. In Sec.~\ref{sec:flat} we show the flat directions allowed by the model. The last section is devoted to our conclusions.

\section{Particle content }
\label{sec:particles}

As we said in the introduction, the gauge symmetry of the model is~\cite{Montero:2007cd}
\begin{equation}
SU(3)_{C}\otimes SU(2)_{L}\otimes U(1)_{Y^\prime}\otimes U(1)_{B-L},  \label{group}
\end{equation}
where $Y^{\prime}$ is a new charge. In this model,
the parameter $Y^\prime$ is chosen to obtain the hypercharge $Y$ of the SM, given by
$Y=Y^\prime+(B-L)$. Thus, in this case, the charge operator is given by
\begin{equation}
\frac{Q}{e}=I_{3}+ \frac{1}{2} \left[ Y^{\prime}+(B-L) \right] .
\end{equation}
We also assume that the $B-L$ and $Y^\prime$ assignments are restricted to  integer numbers.

The particle content of the model is written in terms of superfields. We introduce the leptons by the following left-handed chiral superfields:
\begin{eqnarray}
\hat{L}_{i} &=& \left(
\begin{array}{c}
\hat{\nu}_{i} \\
\hat{l}_{i}
\end{array}
\right) \sim ({\bf 1},{\bf 2},0,-1), \,\
\hat{E}_{i} \sim ({\bf 1},{\bf 1},1,1), \,\
\hat{N}_{i} \sim ({\bf 1},{\bf 1},-1,1),
\label{leptonsm1}
\end{eqnarray}
where we use the notation $\hat{E}_{i}\equiv (\hat{l}_{Ri})^c$ and $\hat{N}_{i}\equiv (\hat{\nu}_{Ri})^c$ for denoting the superfield related with the charge conjugate of the right-handed charged leptons and neutrinos, respectively, with $i=1,2,3$, and in parentheses we present the transformations properties under the respective gauge factors $(SU(3)_{C},SU(2)_{L},U(1)_{Y^{\prime}},U(1)_{B-L})$.
The  chiral superfields that include the known leptons defined above are given in Table~\ref{table1}.

The quarks are also introduced in the following left-handed chiral superfields:
\begin{eqnarray}
\hat{Q}_{i} = \left(
\begin{array}{c}
\hat{u}_{i} \\
\hat{d}_{i}
\end{array}
\right) \sim \left({\bf 3},{\bf 2},0,\frac{1}{3}\right),
\label{quarkssm1}
\end{eqnarray}
and singlets,
\begin{eqnarray}
\hat{U}^c_{i} \sim \left({\bf 3}^*,{\bf 1},-1,-\frac{1}{3}\right),\,\
\hat{D}^c_{i} \sim \left({\bf 3}^*,{\bf 1},1,-\frac{1}{3}\right). \nonumber \\
\label{quarksm2}
\end{eqnarray}
The particle content of these chiral superfields is given in Table~\ref{table2}.

Higgs scalars are also included in left-handed chiral superfields:
\begin{eqnarray}
\hat{H}_{1} &=& \left(
\begin{array}{c}
\hat{h}^{+}_{1} \\
\hat{h}^{0}_{1}
\end{array}
\right) \sim ({\bf 1},{\bf 2},+1,0), \,\
\hat{H}_{2} = \left(
\begin{array}{c}
\hat{h}^{0}_{2} \\
\hat{h}^{-}_{2}
\end{array}
\right) \sim ({\bf 1},{\bf 2}^*,-1,0),
\label{scm1}
\end{eqnarray}
with vacuum expectation values (VEVs) that break the SM symmetry given by
$\langle h^{0}_{1}\rangle = (v_{1}/ \sqrt{2})$ and $\langle h^{0}_{2} \rangle = (v_{2}/ \sqrt{2})$, which will be constrained by the relation $v_1^2+v_2^2 \approx(246\, \textrm{GeV})^2$.
We also add two $SU(2)_{L}$  singlets
\begin{eqnarray}
\hat{\phi}_{1} &\sim& ({\bf 1},{\bf 1},-2,2), \,\
\hat{\phi}_{2} \sim ({\bf 1},{\bf 1},2,-2).
\label{esc1model}
\end{eqnarray}
The second superfield, $\hat{\phi}_{2}$, is necessary in order to cancel the
$U(1)_{Y^\prime}$ and $U(1)_{B-L}$ anomalies produced by the fermionic member of the first one, the $\hat{\phi}_{1}$
superfield. We will denote
the VEVs of the new scalars in the following way: $\langle \phi_{1}\rangle = (u_{1}/ \sqrt{2})$ and
$\langle \phi_{2} \rangle = (u_{2}/ \sqrt{2})$. The values of $u_{1}$ and $u_{2}$ that give the scale of the
$U(1)_{B-L}$ symmetry breaking are not fixed, they may have values ranging from TeV to much higher scales.
See Table~\ref{table3}.

Concerning the gauge bosons and their superpartners, they are introduced in vector superfields. See Table~\ref{table4} the particle content together with the gauge coupling constant of each group.

\section{The Lagrangian}
\label{sec:lagrangian}


Let us begin defining the $R$-parity in the model with the particle content of the previous section.
The connection between R-parity, spin $(S)$, $B$, and  $L$,  can be made explicitly  by writing
\begin{equation}
\hbox{R-parity} = (-1)^{2S} (-1)^{3(B-L)}.
\label{eq:rp02}
\end{equation}
Therefore the  $B-L$ symmetry implies that the $R$-parity is conserved  as a consequence
of the $B-L$ symmetry. The $B-L$ and $R$-parity values of all the fermions of the  model are shown
in Tables~\ref{table5} and \ref{table6}.

An interesting feature of this model is that the neutrinos do not mix with the higgsinos because they have
opposite $R$-parity, and as in the MSSM the lighest supersymmetric particle (LSP) is stable and a possible
candidate to  dark matter.

The  Lagrangian of the model is built with the superfields given in Sec.~\ref{sec:particles} and has the following form
\begin{equation}
{\cal L} = {\cal L}_{\textrm{SUSY}} + {\cal L}_{\textrm{Soft}}, \label{l1}
\end{equation}
where, as usual, ${\cal L}_{\textrm{SUSY}}$ is the supersymmetric piece, while ${\cal L}_{\textrm{Soft}}$ explicitly breaks SUSY.
Below we  write ${\cal L}_{\textrm{SUSY}}$ in terms of the respective superfields, while in Subsec.~\ref{subsec:soft}
we write ${\cal L}_{\textrm{Soft}}$ in terms of the fields.

\subsection{The supersymmetric Lagrangian}
\label{subsec:susyterms}

The supersymmetric term can be divided as follows~\cite{drees,tata}
\begin{equation}
{\cal L}_{\textrm{SUSY}} = {\cal L}_{\textrm{Lepton}} + {\cal L}_{\textrm{Quark}} + {\cal L}_{\textrm{Gauge}} + {\cal L}_{\textrm{Scalar}},
\label{l2}
\end{equation}
we will omit here the ${\cal L}_{\textrm{Quark}}$ term because it is very similar to the MSSM one. The first term in Eq.~(\ref{l2}) is given by
\begin{eqnarray}
{\cal L}_{\textrm{Lepton}}&=&{\cal L}^{\textrm{charged}}_{\textrm{Lepton}}+{\cal L}^{\textrm{neutral}}_{\textrm{Lepton}},  \label{l3}
\end{eqnarray}
where
\begin{eqnarray}
{\cal L}^{\textrm{charged}}_{\textrm{Lepton}}= \int d^{4}\theta\; \sum_{i=1}^{3}\left[\,
\hat{ \bar{L}}_{i}e^{2[g\hat{W}+g_{BL} \left( -\frac{1}{2} \right) \hat{b}_{BL}]}\hat{L}_{i}+
\hat{ \bar{E}}_{i} e^{2[g_{Y^{\prime}} \left( \frac{1}{2} \right)
\hat{b}_{Y^{\prime}}+g_{BL} \left( \frac{1}{2} \right) \hat{b}_{BL}]}\hat{E}_{i} \right].
\label{l3charg}
\end{eqnarray}
In the expressions above we have used $\hat{W}=T^{a}\hat{W}^{a}$ where $T^a=\sigma^{a}/2$ (with $a=1,2,3$) are the
generators of $SU(2)_{L}$ while $g_{Y^{\prime}}$ and $g_{BL}$ are the gauge constant constants of the $U(1)_{Y^{\prime}}$ and the
$U(1)_{B-L}$, respectively, as showed in Table~\ref{table4}. The  gauge coupling constants are related  by the
following relation
\begin{eqnarray}
\frac{1}{g^{2}_{Y}}= \frac{1}{g^2_{Y^\prime}}+ \frac{1}{g^{2}_{BL}},
\label{runcouplingconstant}
\end{eqnarray}
where $g_{Y}$ is the standard model $U(1)_{Y}$ coupling constant.

The second term in Eq.(\ref{l3}) is written as
\begin{eqnarray}
{\cal L}^{\textrm{neutral}}_{\textrm{Lepton}}= \int d^{4}\theta\; \sum_{i=1}^{3}\left[\, \hat{ \bar{N}}_{i}
e^{2[g_{Y^{\prime}} \left( -\frac{1}{2} \right) \hat{b}_{Y^{\prime}}+g_{BL} \left( \frac{1}{2} \right)
\hat{b}_{BL}]} \hat{N}_{i} \,\right].  \label{l3neu}
\end{eqnarray}

The gauge part is given by
\begin{eqnarray}
{\cal L}_{\textrm{Gauge}} &=& \frac{1}{4} \int d^{2}\theta\; \left[ \sum_{A=1}^{8}W^{A}_{c}W^{A}_{c}
+ \sum_{a=1}^{3} W^{a}W^{a}+W^{Y^{\prime}}W^{Y^{\prime}}+W^{BL}W^{BL} \right]
+ H.c. \,\ , \nonumber \\
\label{l5}
\end{eqnarray}
where the strength fields are defined as
\begin{eqnarray}
W^{A}_{\alpha c}&=&- \frac{1}{8g_{s}} \bar{D} \bar{D} e^{-2g_{s} \hat{G}}
D_{\alpha} e^{2g_s \hat{G}},  \,\
W^{a}_{\alpha}=- \frac{1}{8g} \bar{D} \bar{D} e^{-2g \hat{W}} D_{\alpha}
e^{2g \hat{W}},  \nonumber \\
W^{Y^{\prime}}_{\alpha}&=&- \frac{1}{4} \bar{D} \bar{D} D_{\alpha} \hat{b}_{Y^{\prime}},  \,\
W^{BL}_{\alpha}=- \frac{1}{4} \bar{D} \bar{D} D_{\alpha} \hat{b}_{BL} \,\ ,
\label{l6}
\end{eqnarray}
where the superfields are shown in Table IV, and  the covariant derivatives  are given by~\cite{wb}
\begin{equation}
D_{\alpha}(y,\theta,\bar{\theta}) = \frac{\partial}{\partial \theta^{\alpha}}+2i
\sigma^{m}_{\alpha \dot{\alpha}}\bar{\theta}^{\dot{\alpha}}\frac{\partial}{\partial y^{m}} \,,\qquad
\bar{D}_{\dot{\alpha}}(y,\theta,\bar{\theta})=- \frac{\partial}{\partial \bar{\theta}^{\dot{\alpha}}} \,\ .
\label{The Non-Abelian Fieldstrength prop 4}
\end{equation}

Finally, the scalar part in Eq.(\ref{l2}) is
\begin{eqnarray}
{\cal L}_{\textrm{Scalar}} &=& \int d^{4}\theta\;\left[\,
\hat{ \bar{H}}_{1}e^{2[g\hat{W}+g_{Y^{\prime}} \left( \frac{1}{2} \right) \hat{b}_{Y^{\prime}}]}\hat{H}_{1}+
\hat{ \bar{H}}_{2}e^{2[g\hat{W}+g_{Y^{\prime}} \left( \frac{-1}{2}
\right) \hat{b}_{Y^{\prime}}]}\hat{H}_{2}\right.  \nonumber \\
&+& \left. \hat{ \bar{\phi}}_{1}e^{2[g_{Y^{\prime}} \left( \frac{-2}{2}
\right) \hat{b}_{Y^{\prime}}+g_{BL} \left( \frac{2}{2} \right) \hat{b}_{BL}]}\hat{\phi}_{1}+
\hat{ \bar{\phi}}_{2}e^{2[g_{Y^{\prime}} \left( \frac{2}{2}
\right) \hat{b}_{Y^{\prime}}+g_{BL} \left( \frac{-2}{2} \right) \hat{b}_{BL}]}
\hat{\phi}_{2}\right]  \nonumber \\
&+& \left(\int d^2\theta\, W+ H.c.
\right),  \label{l7}
\end{eqnarray}
where $W$ is the superpotential given by
\begin{equation}
W=\frac{W_{2}}{2}+ \frac{W_{3}}{3},  \label{sp1m1}
\end{equation}
where $W_{2}$ and $W_3$  contain  only products of two and three chiral superfields, respectively, and we have written them as
$W_{2,3}=W^{\textrm{MSSM}}_{2,3}+W^{B-L}_{2,3}$.
Explicitly, the terms permitted by the
symmetry have the following form
\begin{eqnarray}
W^{\textrm{MSSM}}_{2}&=&\mu_{H}\hat{H}_{1}\hat{H}_{2}, \,\
W^{B-L}_{2}=\mu_{\phi}\hat{\phi}_{1}\hat{\phi}_{2},  \label{sp2m1}
\end{eqnarray}
and
\begin{eqnarray}
W^{\textrm{MSSM}}_{3}&=&f^l_{ij}\hat{H}_{2}\hat{L}_{i}\hat{E}_{j}, \quad
W^{B-L}_{3}=f^{\nu}_{ij}\hat{H}_{1}\hat{L}_{i}\hat{N}_{j}+f^{M}_{ij}\hat{\phi}_{2}\hat{N}_{i}\hat{N}_{j},   \label{sp3m1}
\end{eqnarray}
where $\hat{H}_{1}\hat{H}_{2}\equiv \epsilon_{\alpha \beta}\hat{H}^{\alpha}_{1}\hat{H}^{\beta}_{2}$.
Again we have omitted summation indices and the quark sector due its similarity with the MSSM.

In general the  parameters $\mu$, in Eq.~(\ref{sp2m1}), and $f^{l,\nu,M}$, in Eq.~(\ref{sp3m1}),  are complex
numbers~\cite{drees,tata}.  The terms proportional to $f^{\nu}$ generate the Dirac mass term $M^{D}=f^{\nu} \langle H_{2}\rangle$, and those proportional to $f^{M}$
generate the Majorana mass term  $M_{M}=f^{M} \langle \phi_{2}\rangle$, which are necessary to
implement  the type I see-saw mechanism  for the active neutrinos~\cite{seesaw}, as we will present at
Subsec.~\ref{subsec:leptonmasses}.

\subsection{Soft Lagrangian}
\label{subsec:soft}

Now we are considering the last source to construct the scalar potential.
The most general soft supersymmetry breaking terms, which do not induce
quadratic divergence, were described by Girardello and Grisaru~\cite{Girardello:1981wz}.
They depend on the model under consideration and in our case they  can be written  as
\begin{eqnarray}
{\cal L}_{\textrm{Soft}} &=& {\cal L}_{\textrm{SMT}} +{\cal L}_{\textrm{GMT}}+{\cal L}_{\textrm{Int}}, \nonumber \\
\label{SoftSUSYm1}
\end{eqnarray}
where the scalar mass term ${\cal L}_{\textrm{SMT}}$ (omitting the squark terms), is given by
\begin{eqnarray}
{\cal L}_{\textrm{SMT}} &=& -\left[\, M_{L}^{2}\;\tilde{L}^{\dagger}\tilde{L} +
M^{2}_{E} \tilde{E}^{\dagger}\tilde{E} +
M^{2}_{N} \tilde{N}^{\dagger}\tilde{N} +
M_{H_{1}}^{2}H^{\dagger}_{1}H_{1}
+M_{H_{2}}^{2} H^{\dagger}_{2}H_{2} +
M_{\phi_{1}}^{2} \phi^{\dagger}_{1}\phi_{1}\right. \nonumber \\ &+& \left.
M_{\phi_{2}}^{2}\phi^{\dagger}_{2}\phi_{2}+
\left( \beta^{2}_{H}H_{1}H_{2}+ \beta^{2}_{\phi} \phi_{1}\phi_{2}+ H.c. \right) \right],
\label{smtsoft}
\end{eqnarray}
where the scalar mass terms $M_{L}^{2}$, $M^{2}_{N}$ and $M_{E}^{2}$ are, in general, hermitian
$3\!\times\!3$ matrices in the generation space~\cite{drees,tata}, and $\beta_{\phi,H}$ are parameters with dimension of mass.

The gaugino mass term ${\cal L}_{\textrm{GMT}}$ is defined as
\begin{eqnarray}
{\cal L}_{\textrm{GMT}} &=&-\frac{1}{2} \left(\,M_{\tilde{g}}\; \sum_{A=1}^{8}
\tilde{g}^{A} \tilde{g}^{A} +
M_{\tilde{W}}\; \sum_{a=1}^{3}\;\tilde{W}^{i} \tilde{W}^{i}+
M_{\tilde{b}_{Y^{\prime}}} \; \tilde{b}_{Y^{\prime}} \tilde{b}_{Y^{\prime}} +
M_{\tilde{b}_{BL}}\; \tilde{b}_{BL} \tilde{b}_{BL} \,\right) +
H.c., \nonumber \\
\end{eqnarray}
and the last term ${\cal L}_{\textrm{Int}}$ is
\begin{eqnarray}
{\cal L}_{\textrm{Int}} &=& A_{l} H_{2}\tilde{L}\tilde{E}+
A_{d} H_{2}\tilde{Q}\tilde{D}+ A_{u} H_{1}\tilde{Q}\tilde{U}+
A_{\nu} H_{1}\tilde{L}\tilde{N} + A_{N}\phi_{2}\tilde{N}\tilde{N}+H.c.. \nonumber \\
\label{burro}
\end{eqnarray}

Finally, the gauge symmetry breaking pattern of the model is given by
\begin{equation}
 \mbox{SU(2)}_{L}\otimes \mbox{U(1)}_{Y^{\prime}}\otimes \mbox{U(1)}_{B-L}
\stackrel{\langle\phi_{1}\rangle , \langle \phi_{2}\rangle}{\longrightarrow}
 \mbox{SU(2)}_{L}\otimes \mbox{U(1)}_{Y}
 \stackrel{\langle H_{1}\rangle , \langle H_{2}\rangle}{\longrightarrow}
 \mbox{U(1)}_{Q}.
\end{equation}
Note that only after the first spontaneous symmetry breakdown, when $\langle\phi_{1}\rangle \neq 0$ and
$\langle \phi_{2}\rangle \neq 0$, we obtain the usual SM symmetries. Taking this fact into account, we can consider the
hierarchy $u_{1},u_{2} \gg v_{1},v_{2}$, between the VEVs of this model.

\section{Gauge boson and fermion mass spectra}
\label{sec:masses}

We have discussed the particle content as symmetry eigenstates, now we will show the mass eigenstates
of the model. Once the $SU(2)_{L} \otimes U(1)_{Y^{\prime}}\otimes U(1)_{B-L}$ symmetry is broken, fields
with the same $SU(3)_{C} \otimes U(1)_Q$ quantum numbers can mix with each other.  For instance, the Dirac masses of quarks and leptons can be understood as such mixing terms. In the MSSM, this mixing also affects
squarks, sleptons, Higgs bosons, as well as gauginos and higgsinos. The only
exception is the gluino, being the only color  fermion octet in the model~\cite{drees,tata}. In this model the gluinos are
the same as in the MSSM and, hence, we will not reproduce them in this article.

\subsection{The masses of the gauge vector bosons}
\label{subsec:vectors}

The charged gauge vector boson mass is given by
\begin{equation}
M^{2}_{W}=\frac{g^{2}v^{2}_{2}}{4} \left( 1+ \tan^{2} \beta \right) \,\ ,
\label{wmass}
\end{equation}
where we have defined $\beta$ through the ratio
\begin{equation}
\tan \beta \equiv \frac{v_{1}}{v_{2}},\qquad 0 \leq \beta \leq \frac{\pi}{2}\, \textrm{rad}.
\label{tanb}
\end{equation}

\begin{figure}[h]
\centering
\includegraphics[width=5in]{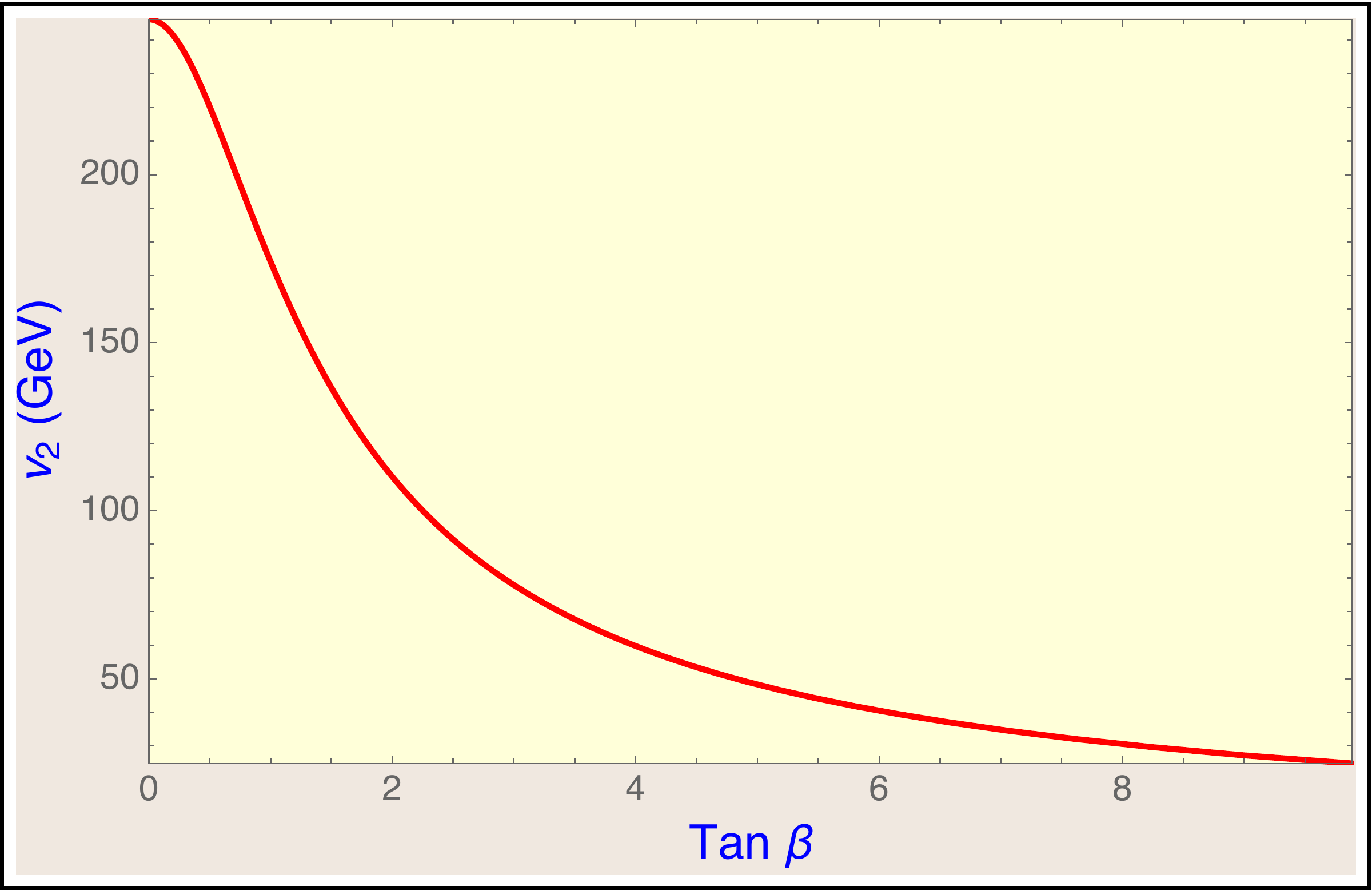}
\caption{The values of $v_{2}$ in terms of $\tan \beta$ in agreement with $M_{W}=80.363$ GeV using Eq.~(\ref{wmass}).}
\label{fig1}
\end{figure}

In this work, we use the following values for the constant couplings: $g=0.653$, the same expression used in the SM, $g_{Y^\prime}=0.485$, and
$g_{BL}=0.506$, using Eq.~(\ref{runcouplingconstant}). Our first numerical result is presented in Fig.~\ref{fig1}, where we
impose the experimental value of the $W$ boson mass, $M_{W}=80.363$ GeV, and we find the value of $v_{2}$
in terms of $\tan \beta$ that satisfy this experimental limit using Eq.(\ref{wmass}).

The square mass matrix for the neutral vector bosons in the
$(W^{3},b_{Y^{\prime}}$, $b_{BL})$ base is
\begin{eqnarray}
M^{2}_{{\rm neutral}}= \left(
\begin{array}{ccc}
\frac{g^{2}}{4}(v_{1}^{2}+v_{2}^{2}) & -
\frac{gg_{Y^{\prime}}}{4}(v_{1}^{2}+v_{2}^{2}) & 0 \\
- \frac{gg_{Y^{\prime}}}{4}(v_{1}^{2}+v_{2}^{2}) & g^{2}_{Y^{\prime}} \left(
\frac{v_{1}^{2}+v_{2}^{2}}{4}+u_{1}^{2}+u_{2}^{2} \right) & -
g_{Y^{\prime}}g_{BL}(u_{1}^{2}+u_{2}^{2}) \\
0 & -g_{Y^{\prime}}g_{BL}(u_{1}^{2}+u_{2}^{2}) &
g^{2}_{BL}(u_{1}^{2}+u_{2}^{2})
\end{array}
\right), \nonumber \\
\end{eqnarray}
with ${\rm det}\,M^2_{{\rm neutral}}=0$. The exact mass eigenvalues, as given
in Ref.~\cite{Fortes:2010mz}, are: zero for the photon field, and two massive
fields with masses given by
\begin{eqnarray}
M^{2}_{1,2}=\frac{1}{8}\left(U \mp \sqrt{U^{2}-V} \right),
\label{massaapprox1}
\end{eqnarray}
where
\begin{eqnarray}
U&=&4\left(g_{Y^{\prime}}^{2}+g_{BL}^{2}\right)\left(u_{1}^{2}+u_{2}^{2}\right)
+\left(g^{2}+g_{Y^{\prime}}^{2}\right)\left(v_{1}^{2}+v_{2}^{2}\right),  \nonumber \\
V&=&16\left[g^{2}\left(g_{Y^{\prime}}^{2}+g_{BL}^{2}\right)+g_{Y^{\prime}}^{2}g_{BL}^{2}\right]
\left(u_{1}^{2}+u_{2}^{2}\right)\left(v_{1}^{2}+v_{2}^{2}\right).
\label{massaapprox2}
\end{eqnarray}
Using Eq.(\ref{wmass}) we can rewrite the previous expressions as
\begin{eqnarray}
U&=&4u_{1}^{2}\left(g_{Y^{\prime}}^{2}+g_{BL}^{2}\right) \left(1+ \tan^{2} \xi \right)
+ \frac{4M^{2}_{W}}{g^{2}}\left(g^{2}+g_{Y^{\prime}}^{2}\right),  \nonumber \\
V&=&\frac{64M^{2}_{W}u_{2}^{2}}{g^{2}}\left[g^{2}\left(g_{Y^{\prime}}^{2}+g_{BL}^{2}\right)+g_{Y^{\prime}}^{2}g_{BL}^{2}\right]
 \left(1+ \tan^{2} \xi \right),
\label{massaapprox4}
\end{eqnarray}
with the definition
\begin{equation}
\tan \xi \equiv \frac{u_{1}}{u_{2}},\qquad 0 \leq \xi \leq  \frac{\pi}{2}\,\, \textrm{rad}.
\label{tanXi}
\end{equation}

We can reproduce the values of the non supersymmetric model, presented in Ref.~\cite{Montero:2007cd}, when we consider $\beta = \xi =0$ rad, and
if we want to reproduce the experimental
data for the neutral vector boson mass, $M_{Z}=91.1876$ GeV, we need to use $u_{2}=5$ TeV. In this case we get $M_{Z^{\prime}}=3.5$ TeV, and the
experimental limit is $M_{Z^{\prime}}>3$ TeV. We see that the numerical value to the $Z^{\prime}$ gauge boson mass also
satisfies the relation $M_{Z^{\prime}}/g_{BL}=6.9284>6$ TeV \cite{Carena:2004xs}.

\begin{figure}[h]
\centering
\includegraphics[width=5in]{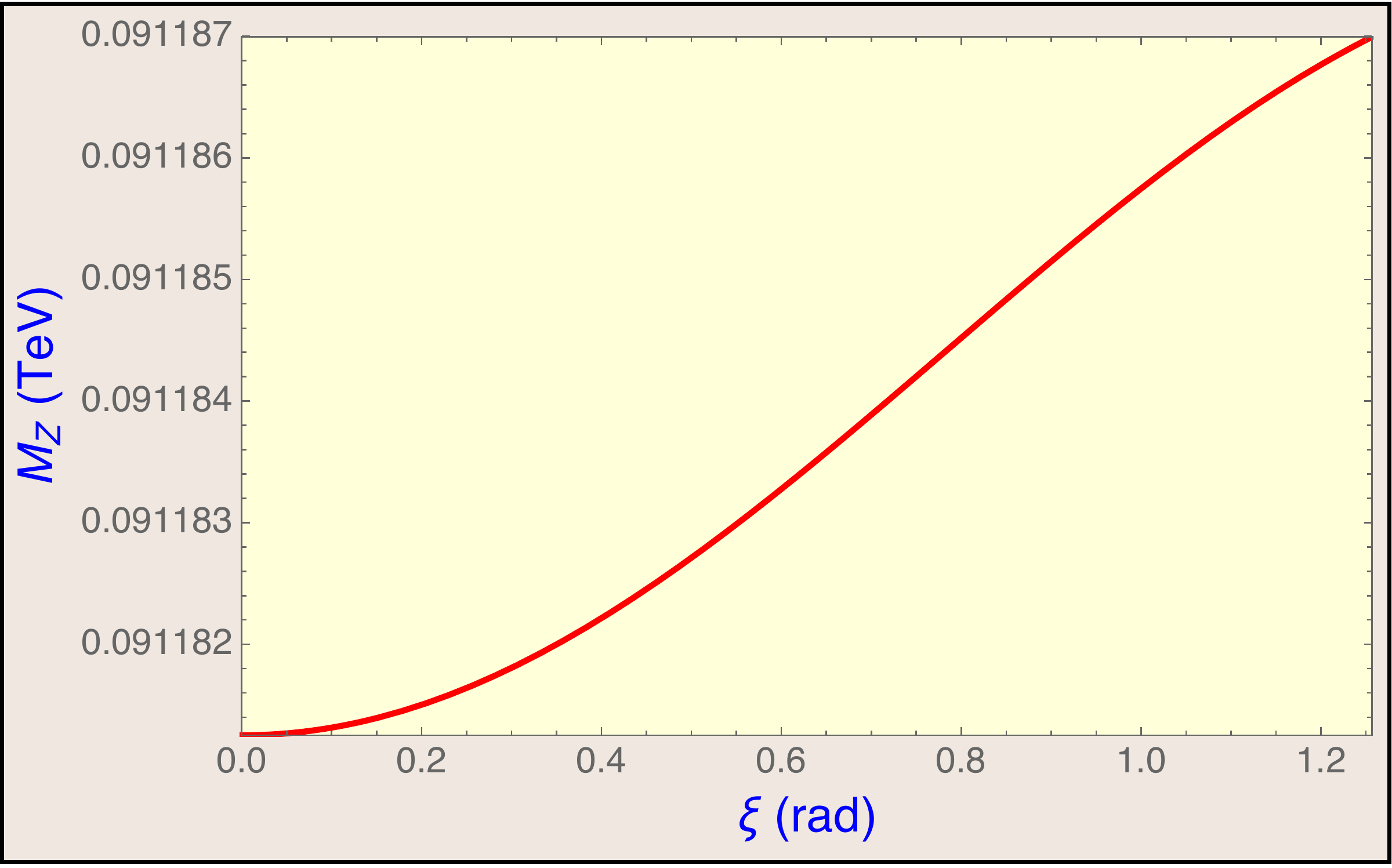}
\caption{The values of $M_{Z}=M_{1}$ in terms of $\tan \xi$  using Eq.(\ref{massaapprox1}) and (\ref{massaapprox4}).}
\label{fig2}
\end{figure}

\begin{figure}[h]
	\centering
	\includegraphics[width=5in]{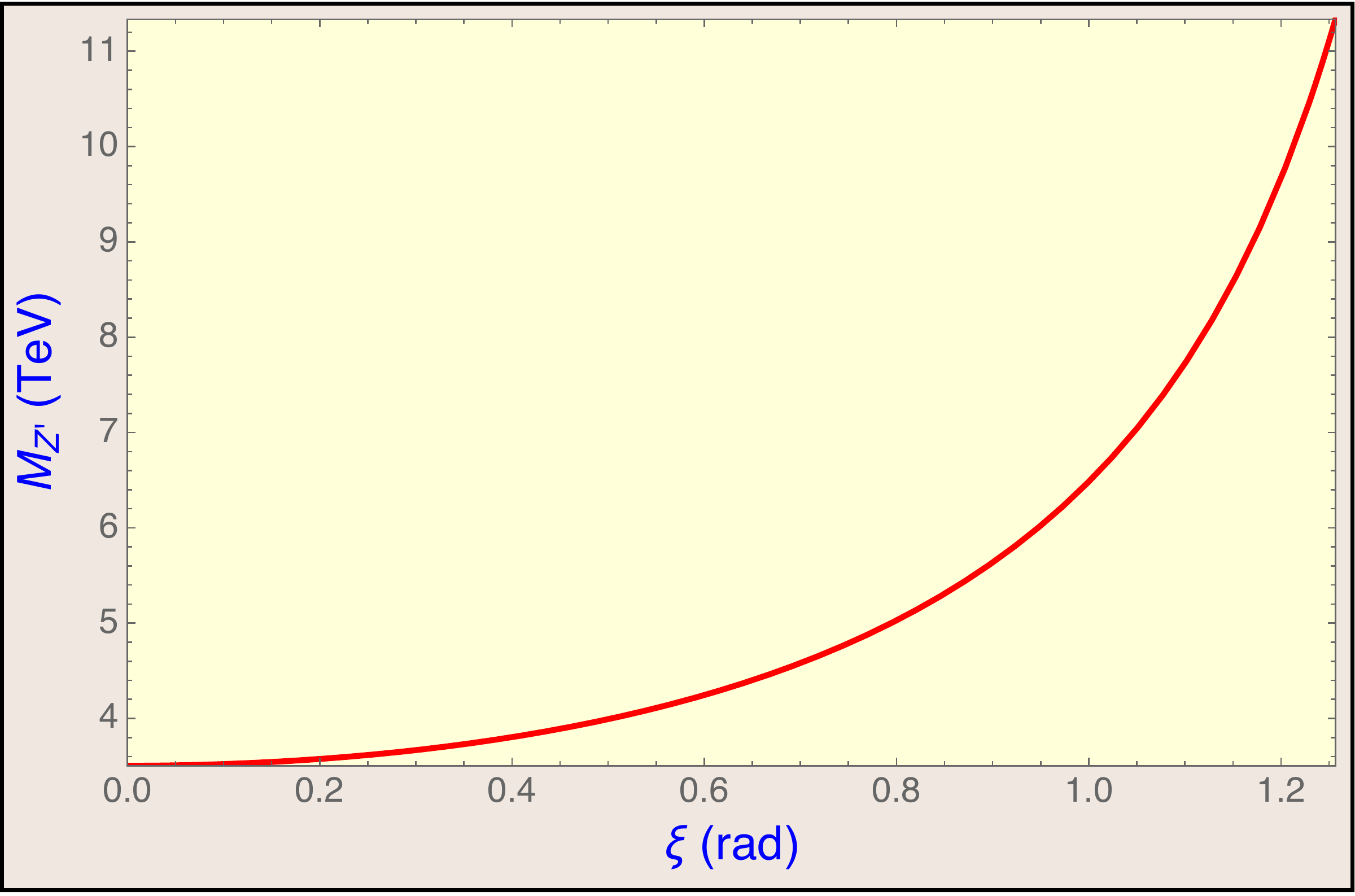}
	\caption{The values of $M_{Z^\prime}=M_{2}$ in terms of $\tan \xi$  using Eq.(\ref{massaapprox1}) and (\ref{massaapprox4}).}
	\label{fig3}
\end{figure}

However, when we plot $M_{Z}$, we can reproduce the experimental data using any value of $\xi$ as we
can see in Fig.~\ref{fig2}. Moreover, $M_{Z^{\prime}}$ is bigger than $3$ TeV when
$0 < \xi < (\pi /2)$ rad as we shown in Fig.~\ref{fig3}. The neutral gauge  boson sector is the same as presented
in Refs.~\cite{Fortes:2010mz,Coutinho:2011xb,Bruce:2014DM}.

\subsection{Lepton masses}
\label{subsec:leptonmasses}

Now, we will calculate the mass spectrum from the charged leptons, the mechanism to give mass to the quarks is similar to
that of the MSSM, at the tree level. The  mass source for all the fermions in supersymmetric models comes
from the superpotencial $W_{3}$. The Yukawa interactions in the lepton sector are given by
\begin{eqnarray}
{\cal L}^{W_3}_{ffH}=- \frac{1}{3} \sum_{i,j=1}^{3}\left(f^{l}_{ij}\epsilon_{\alpha \beta}L^\alpha_{i}E_{j}H^\beta_{2}+
f^{\nu}_{ij}\epsilon_{\alpha \beta}L^{\alpha}_{i}N_{j}H^{\beta}_{1}+
f^{M}_{ij}\phi_{2}N_{i}N_{j}\right) +H.c .
\label{fermionsm1}
\end{eqnarray}
It is straightforward  to show that the charged lepton mass matrix can be written as 
\begin{eqnarray}
M^{l}_{ij}=\frac{\sqrt{2}M_{W} \cos \beta}{g}f^{l}_{ij}.
\label{chargedlepton}
\end{eqnarray}


In the basis $\Psi^{0}=(\nu_{1} ,\, \nu_{2},\,\nu_{3} ,\, N_{1} ,\, N_{2}, \, N_{3})^T$, the neutrino mass term   is writing as
$-(1/2) \Psi^{0T}Y^{0}\Psi^{0}+H.c.$, where $Y^{0}$ is the mass matrix given by
\begin{equation}
\left(
\begin{array}{cc}
0_{3 \times 3} & v_{2}\tan \beta f^{\nu} \\
v_{2}\tan \beta f^{\nu T} & \frac{u_{2}}{\sqrt{2}}f^{M}
\end{array}
\right),
\end{equation}
and we see that it can generate the type I seesaw mechanism \cite{seesaw}. In this context, it is usual to introduce the notation: $M^{D}=v_{2}\tan \beta f^{\nu}$ and $M^{M}=\frac{u_{2}}{\sqrt{2}}f^{M}$. It leads to the following expressions for
light and heavy neutrino masses, respectively
\begin{equation}
M^{\nu}_{L}\approx-M^{D}\left( M^{M} \right)^{-1} M^{D T}, \quad
M^{\nu}_{H}\approx M^{M}.
\label{eq:seesawtoneutrinos}
\end{equation}
Since we are implementing the type-I seesaw mechanism we do not need to impose severe constraints  in order to explain the
neutrino masses. We can accommodate the experimental data if $f^{\nu}\sim{\cal O}(10^{-4})$. As we will show below, the model has flat directions, therefore, it is an example of a supersymmetric model in which the problems of the smallness of neutrino masses, inflation, dark matter and baryon asymmetry will be related, without the strong constraint, $\vert f^{\nu}_{ij}\vert{\lesssim}10^{-12}$, needed  in Ref.~\cite{Allahverdi:2007wt}.

Next,  we will consider charginos and neutralinos which are mass eigenstates of gaugino-higgsino combinations.

\subsection{Chargino masses}
\label{subsec:charginos}

The mass terms in the chargino sector are given by
\begin{equation}
- \frac{\mu_{H}}{2} \tilde{h}^{+}_{1}\tilde{h}^{-}_{2}+M_{\tilde{W}}\; \tilde {W}^{+}\tilde{W}^{-}-
ig \left( v_{2}\tilde {W}^{+}\tilde{h}^{-}_{2}+v_{1}\tilde {W}^{-}\tilde{h}^{+}_{1} \right) +H.c. \,\ ,
\end{equation}
where $\tilde{W}^{\pm}=(\tilde{W}^{1}\mp i \tilde{W}^{2})/\sqrt{2}$,
 as in the MSSM. Using the same parameters of Ref.~\cite{Montero:2001ch}, $\mu_{H}=200$ GeV,
$M_{\tilde{W}}=350$ GeV, and $\tan \beta =1$, we get
\begin{eqnarray}
m_{\tilde{\chi}_{1}}=165.076, \,\
m_{\tilde{\chi}_{2}}=384.924,
\label{charginomasses}
\end{eqnarray}
both values are in GeV, which are in agreement with the experimental limit $m_{\tilde{\chi}_{1}}>45$ GeV~\cite{Agashe:2014kda}.

\subsection{Neutralino masses}
\label{subsec:neutralinos}

In the basis $\Psi^{0}_{N}=(-i \tilde{W}^{3}, -i \tilde{b}_{Y^{\prime}}, -i  \tilde{b}_{BL}, \tilde{h}^{0}_{2} ,
\tilde{h}^{0}_{1},\tilde{\phi}_{1}, \tilde{\phi}_{2})^T$,
the mass term of the neutralinos is written in the following way
\begin{equation}
- \left( \frac{1}{2} \right) \left[ \Psi^{0T}_{N}Y^{0}_{N}\Psi^{0}_{N}+H.c. \right] \,\ ,
\end{equation}
where $Y^{0}_{N}$ is the mass matrix given by
\begin{eqnarray}
Y^{0}_{N}= \left(
\begin{array}{cc}
M_{G} & M_{GH} \\
M^{T}_{GH} & M_{H}
\end{array}
\right). \nonumber \\
\end{eqnarray}
where
\begin{eqnarray}
M_{G}&=& \left(
\begin{array}{cccc}
M_{\tilde{W}} & 0 & 0  \\
0 & M_{\tilde{b}_{Y^{\prime}}} & 0 \\
0 & 0 & M_{\tilde{b}_{BL}}
\end{array}
\right), \,\
M_{H}= \left(
\begin{array}{cccc}
0 &\mu_{H} & 0 & 0  \\
\mu_{H} & 0 & 0 & 0 \\
0 & 0 & 0 &\mu_{\phi} \\
0 & 0 &\mu_{\phi} & 0
\end{array}
\right), \nonumber \\
M_{GH}&=& \left(
\begin{array}{cccc}
M_{W}\sin \beta &-M_{W}\cos \beta & 0 & 0 \\
 \frac{g_{Y^{\prime}}}{g}M_{W}\sin \beta &
\frac{-g_{Y^{\prime}}}{g}M_{W}\cos \beta & g_{Y^{\prime}}u_{2}\tan \xi &- g_{Y^{\prime}}u_{2} \\
0 & 0 &g_{BL}u_{2}\tan \xi &- g_{BL}u_{2}
\end{array}
\right).
\end{eqnarray}
If we choose, besides the parameter defined in Eq.(\ref{charginomasses}), as we have done in Ref.~\cite{Montero:2001ch},  $\mu_{\phi}=100$ GeV, $M_{\tilde{b}_{Y^{\prime}}}=M_{\tilde{b}_{BL}}=-200$ GeV, $u_{2}=5$ TeV, and $\tan \xi =1$, we get
\begin{eqnarray}
m_{\tilde{\chi}_{1}}&=&100, \,\
m_{\tilde{\chi}_{2}}=200, \,\
m_{\tilde{\chi}_{3}}=-215.633, \,\
m_{\tilde{\chi}_{4}}=227.813, \nonumber \\
m_{\tilde{\chi}_{5}}&=&361.778, \,\
m_{\tilde{\chi}_{6}}=-4894.43, \,\
m_{\tilde{\chi}_{7}}=5020.48, \,\
\end{eqnarray}
in GeV. We must emphasize that the lightest neutralino, as it happens in the MSSM, has mass of the order $O(M_{Z})$. This value is in agreement with the experimental limit $m_{\tilde{\chi}_{1}}>30$ GeV~\cite{Agashe:2014kda}.

\section{The scalar potential}
\label{sec:potential}

The scalar potential of this model has the following form
\begin{equation}
V=\sum_{f}F_{f}^{\dagger }F_{f}+\frac{1}{2}\left(
D_{Y^{\prime}}^{2}+D_{BL}^{2}+D^{i}D^{i}+D^{a}D^{a}\right)+V_{\textrm{Soft}},
\end{equation}
where $f=H_{1},H_{2},\phi _{1},\phi _{2},L,E,N$; $i=1,2,3$; and $a=1,...,8$.
The $F$ terms are obtained from the superpotential, and they are
\begin{eqnarray}
F^{\dagger a}_{H_{1}}&=&-\frac{\mu_{H}}{2}H_{2}^{a}-
\frac{f_{ij}^{\nu}}{3}\tilde{L}_{i}^{a}\tilde{N}_{j}, \nonumber\\
F^{\dagger a}_{H_{2}}&=&-\frac{\mu_{H}}{2}H_{1}^{a}-\frac{f_{ij}^{l}}{3}\tilde{L}_{i}^{a}\tilde{E}_{jL},  \nonumber \\
F^{\dagger}_{\phi_{1}}&=&-\frac{\mu_{\phi}}{2}\phi_{2}, \nonumber \\
F^{\dagger}_{\phi_{2}}&=&-\frac{\mu_{\phi}}{2}\phi_{1}-\frac{f_{ij}^{M}}{3}\tilde{N}_{i}\tilde{N}_{j}, \nonumber \\
F^{\dagger a}_{L_{i}}&=&- \frac{f_{ij}^{l}}{3}H_{2}^{a}\tilde{E}_{j}-
\frac{f_{ij}^{\nu}}{3}H_{1}^{a}\tilde{N}_{j}, \nonumber\\
F^{\dagger}_{E_{i}}&=&-\frac{f_{ji}^{l}}{3}H^{a}_{2}\epsilon_{a b}\tilde{L}^{b}_{j}, \nonumber \\
F^{\dagger}_{N_{i}}&=&-\frac{f_{ji}^{\nu}}{3}H^{a}_{1}\epsilon_{ab}\tilde{L}^{b}_{j}-2
\frac{f_{ij}^{M}}{3}\phi_{2}\tilde{N}_{j}.
\label{fterms}
\end{eqnarray}
Where we have omitted the squark terms. There is one $D$-term for each of the four gauge groups
\begin{eqnarray}
&& D_{Y^\prime}=-\frac{1}{2}g_{Y^{\prime}}\left[H_{1}^{\dagger}H_{1}-H_{2}^{\dagger}H_{2}-2\phi_{1}^{\dagger}\phi_{1}+2\phi_{2}^{\dagger}\phi_{2}+\tilde{E}_{i}^{\dagger}\tilde{E}_{i}- \tilde{N}_{i}^{\dagger}\tilde{N}_{i}\right], \nonumber \\ &&
D_{B-L}=-\frac{1}{2}g_{BL}\left[2\phi_{1}^{\dagger}\phi_{1}-2\phi_{2}^{\dagger}\phi_{2}-
\tilde{L}_{i}^{\dagger}\tilde{L}_{i}+\tilde{E}_{i}^{\dagger}\tilde{E}_{i}+
\tilde{N}_{i}^{\dagger}\tilde{N}_{i}\right], \nonumber \\ &&
D^{a}=-\frac{g}{2} \left[H_{1}^{\dagger}\sigma^{a}H_{1}+
H_{2}^{\dagger}\sigma^{a}H_{2}+\tilde{L}_{i}^{\dagger}\sigma^{a}\tilde{L}_{i} \right].
\label{dterms}
\end{eqnarray}


\section{Higgs scalars}
\label{sec:higgs}

Here, we restrict ourselves to calculate only the Higgs potential, $V_{\textrm{Higgs}}$. This is given by
\begin{eqnarray}  \label{ess}
V_{\textrm{Higgs}}&=&V_{\textrm{SUSY}}+V_{\textrm{Soft}}, \nonumber \\
V_{\textrm{SUSY}}&=&|\mu_{H_{1}}|^{2}H_{1}^{\dagger}H_{1}+|\mu_{H_{2}}|^{2}H_{2}^{\dagger}H_{2}+|\mu_{\phi_{2}}|^{2}\phi_{1}^{\dagger}\phi_{1}+|\mu_{\phi_{2}}|^{2}\phi_{2}^{\dagger}\phi_{2}\nonumber\\
&+&\frac{g^{2}}{8} \left[H_{1}^{\dagger}\sigma^{a}H_{1}+H_{2}^{\dagger}\sigma^{a}H_{2} \right]^{2}+ \frac{g_{BL}^{2}}{8}\left[2\phi_{1}^{\dagger}\phi_{1}-2\phi_{2}^{\dagger}\phi_{2}\right]^{2} \nonumber \\ &+&
\frac{g_{Y^{\prime}}^{2}}{8}\left[H_{1}^{\dagger}H_{1}-H_{2}^{\dagger}H_{2}-2\phi_{1}^{\dagger}\phi_{1}+2\phi_{2}^{\dagger}\phi_{2}\right]^{2}, \nonumber \\
V_{\textrm{Soft}}&=& M^{2}_{H_1}H_{1}^{\dagger}H_{1}+ M^{2}_{H_2}H_{2}^{\dagger}H_{2}+ M^{2}_{\phi_1}\phi_{1}^{\dagger}\phi_{1}+
M^{2}_{\phi_2}\phi_{2}^{\dagger}\phi_{2}+
\beta^{2}_{H}\epsilon_{\alpha \beta} H_{1}^{\alpha}H_{2}^{\beta}\nonumber \\ &+&\beta^{2}_{\phi}\phi_{1}\phi_{2}+H.c. .
\label{potential}
\end{eqnarray}
To calculate the mass matrices we make a shift in the neutral scalar fields as
\begin{eqnarray}
H_{1}&=& \left(
\begin{array}{c}
H_{1}^{+} \\
\frac{1}{\sqrt{2}} \left (v_{1}+ \textrm{Re}{H}_{1}^{0}+i \textrm{Im} {H}_{1}^{0} \right)%
\end{array}
\right), \,\
\phi_{1}=\frac{1}{\sqrt{2}} \left(u_{1}+\textrm{Re}{\phi}_{1}+i\textrm{Im}{\phi}_{1} \right),  \nonumber \\
H_{2} &=& \left(
\begin{array}{c}
\frac{1}{\sqrt{2}} \left(v_{2}+ \textrm{Re}{H}_{2}^{0}+i \textrm{Im} {H}_{2}^{0} \right) \\
H_{2}^{-}
\end{array}
\right), \,\
\phi_{2}= \frac{1}{\sqrt{2}} \left(u_{2}+\textrm{Re}{\phi}_{2}+i\textrm{Im}{\phi}_{2}\right).
\end{eqnarray}

\subsection{Charged scalars}
\label{subsec:charged}

The charged scalar  mass matrix, in the basis $(H^{+}_{1},H^{+}_{2})$, is given by
\begin{eqnarray}
M^{2}_{C}=
\left(
\begin{array}{ll}
\frac{g^{2} v_{2}^{2}}{4}+\beta_{H}^{2}\cot \beta  &
\frac{g^{2}v_{1}v_{2}}{4}+\beta_{H}^{2}\tan\beta \\
\frac{g^{2}v_{1}v_{2}}{4}+\beta_{H}^{2}\cot\beta &
\frac{g^{2}v_{1}^{2}}{4}+\beta_{H}^{2}\tan \beta
\end{array}
\right),
\end{eqnarray}
and the determinant of this matrix is vanishing.

The mass spectrum results in one Goldstone, $G^{+}$, which will be absorbed to form the longitudinal component of the charged massive
vector boson $W^{+}$, and one physical mass
eigenstate $h^{+}$, whose mass is given by
\begin{equation}
m_{h^{+}}^{2}=\left( \tan \beta + \cot \beta \right) \beta_{H}^{2}+M^{2}_{W}.
\label{masschargedhiggs}
\end{equation}
In the last passage we have used Eq.(\ref{wmass}) and Eq.(\ref{tanb}). In the MSSM there is one charged Higgs with  the
same mass expression as  derived above~\cite{drees,tata}. The mass values for this charged scalar in the model satisfy
the  relation $200< m_{h^{\pm}}<2000$ GeV~\cite{Aaboud:2016dig}.

The corresponding eigenvectors are given by
\begin{eqnarray}  \label{evchargedscalar}
G^{+}&=&\frac{1}{\sqrt{1+ \tan^{2} \beta}}\left( - \tan \beta H^{+}_{1}+H^{+}_{2} \right), \,\
h^{+}=\frac{1}{\sqrt{1+ \cot^{2} \beta}}\left(\frac{1}{\tan \beta}H^{+}_{1}+H^{+}_{2}\right). \nonumber \\
\end{eqnarray}

\subsection{Pseudoscalars}
\label{subsec:pseudo}

The imaginary squared mass matrix in the basis $(\textrm{Im} {H}_{1}^{0},\textrm{Im} {H}_{2}^{0},\textrm{Im}{\phi}_{1},\textrm{Im}{\phi}_{2})$ is given by
\begin{eqnarray}
M^2_P &=& \left(
\begin{array}{cccc}
\beta_{H}^{2}\cot \beta  & \beta_{H}^{2} & 0 & 0 \\
\beta_{H}^{2} & \beta_{H}^{2}\tan \beta  & 0 & 0 \\
0 & 0 &\beta_{\phi}^{2} \cot \xi  & \beta_{\phi}^{2} \\
0 & 0 & \beta_{\phi}^{2} & \beta_{\phi}^{2}\tan \xi
\end{array}
\right).
\end{eqnarray}
Notice that the basis $(\textrm{Im} {H}_{1}^{0},\textrm{Im} {H}_{2}^{0})$ does not mix with $(\textrm{Im}{\phi}_{1},\textrm{Im}{\phi}_{2})$ and that the determinant of each sub-matrix vanishes. In fact, this mass matrix has two zero eigenvalues corresponding to the two Goldstone bosons, $G_{1},\, G_{2}$
(they will become the longitudinal components of the $Z^{0}$ and $Z^{\prime 0}$ neutral vector bosons), and two mass eigenstates, $A_{1},\, A_{2}$, and we have
defined $A_{2}$ to be the heavier of the two, i. e., $m_{A_{2}}>m_{A_{1}}$. Their masses are given by
\begin{eqnarray}
m_{A_{1}}^{2}&=& \left( \tan \beta + \cot \beta \right) \beta_{H}^{2}, \quad
m_{A_{2}}^{2}= \left( \tan \xi + \cot \xi \right) \beta_{\phi}^{2},
\label{cpimparhiggs}
\end{eqnarray}
where $\tan \beta$ and $\tan \xi$ are defined in Eq.~(\ref{tanb}) and Eq.~(\ref{tanXi}), respectively, and
$\beta_{H},\beta_{\phi}$ are the parameter with dimension of mass in Eq.~(\ref{potential}). The $C\!P$
odd Higgs in the MSSM is the same as  $A_{1}$. Using Eq.(\ref{masschargedhiggs}) we can write, as in the MSSM, the
relation
\begin{eqnarray}
m_{h^{+}}^{2}=m_{A_{1}}^{2}+M^{2}_{W}.
\label{ahp}
\end{eqnarray}

\begin{figure}[h]
	\centering
	\includegraphics[width=5in]{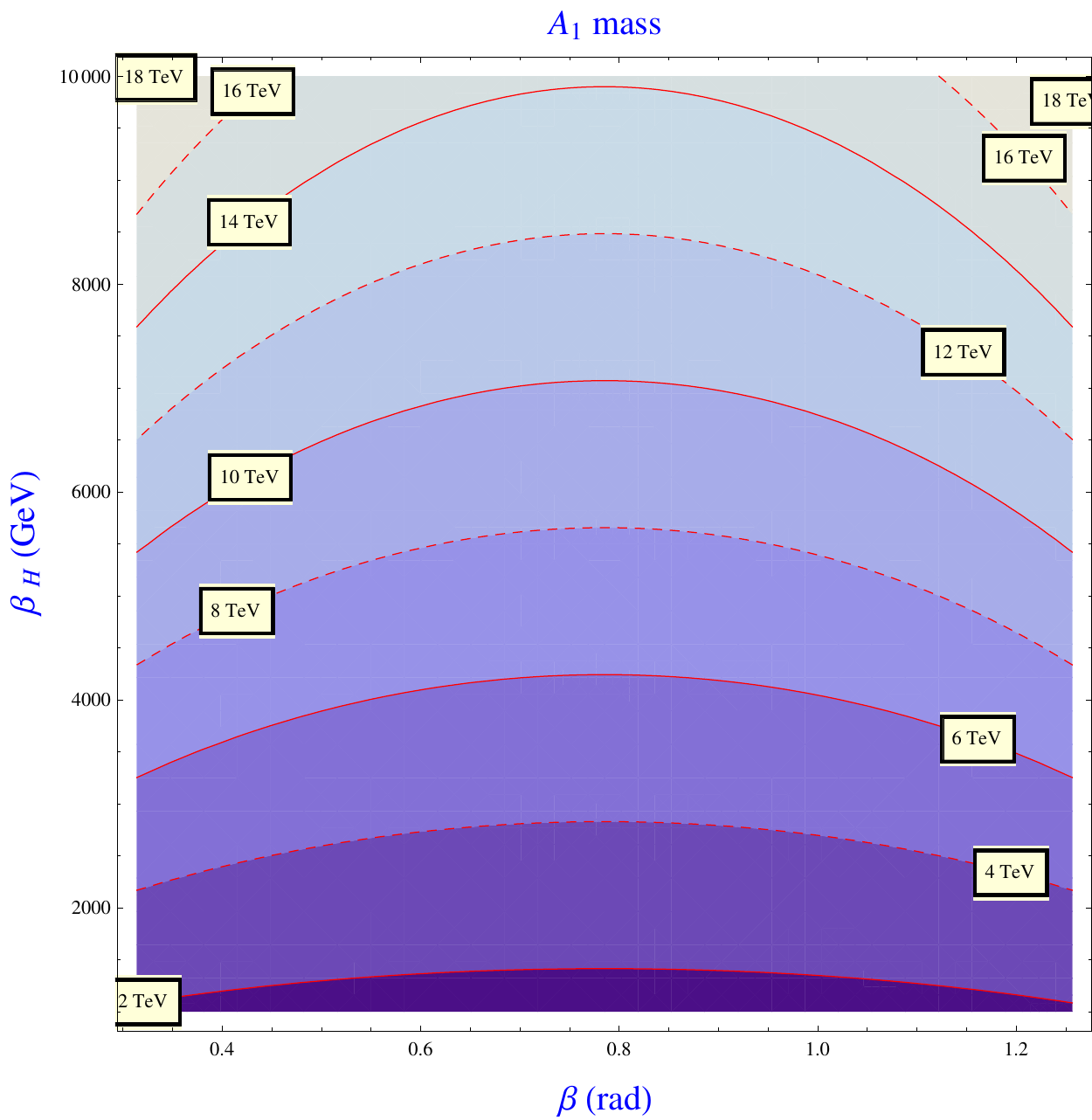}
	\caption{The values of $M_{A_{1}}$ in terms of $\beta$ and $\beta_{H}$ using the first relation in Eq.(\ref{cpimparhiggs}).}
	\label{fig4}
\end{figure}

\begin{figure}[h]
	\centering
	\includegraphics[width=5in]{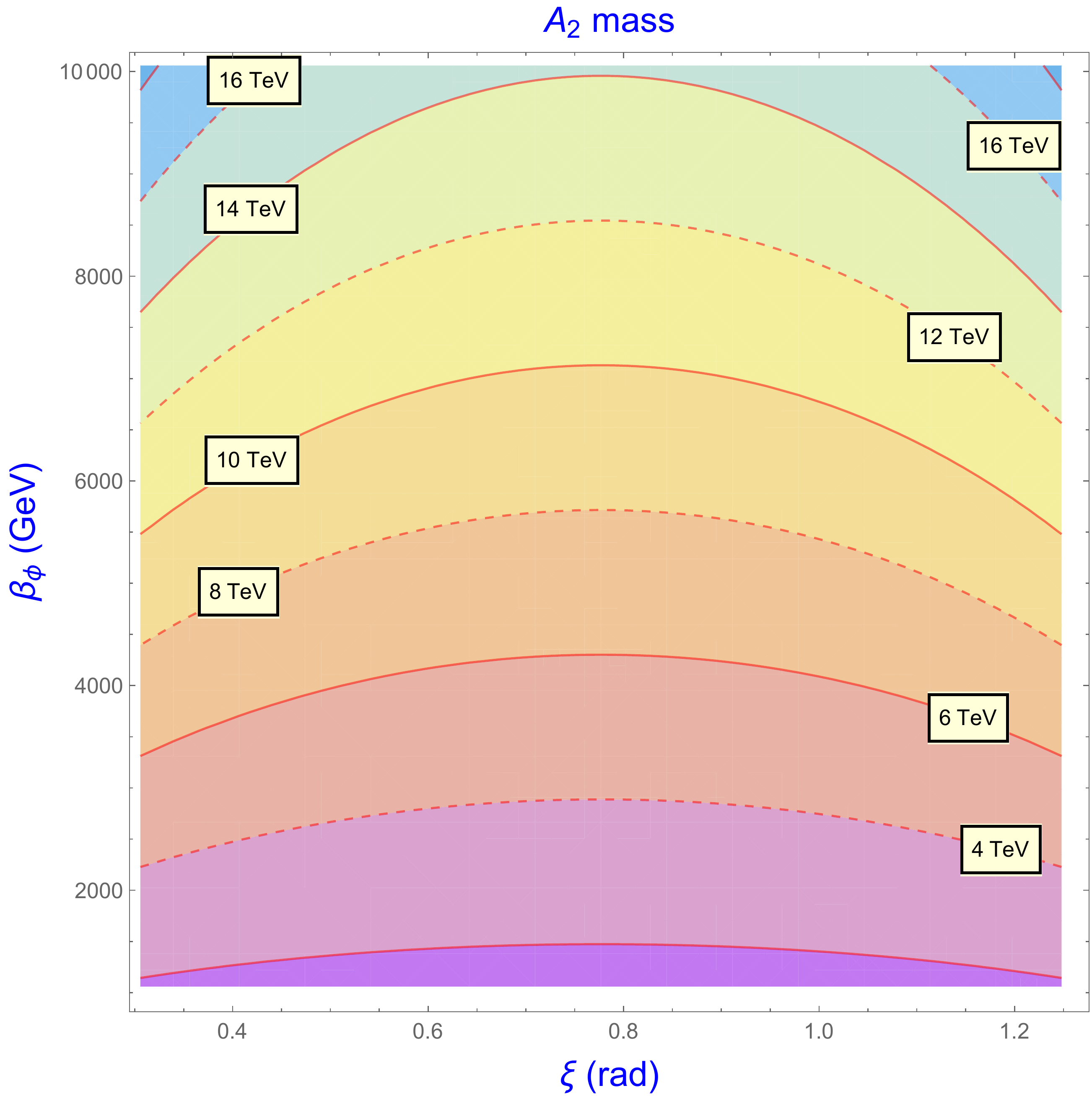}
	\caption{The values of $M_{A_{2}}$ in terms of $\xi$ and $\beta_{\phi}$ using the second relation in Eq.(\ref{cpimparhiggs}).}
	\label{fig5}
\end{figure}

We show the behaviour of the mass of $A_{1}$ in terms of  $\beta$ and the soft parameter $\beta_{H}$ in Fig.~\ref{fig4}. In
Fig.~\ref{fig5} we show the mass of $A_{2}$ as a function of $\Xi$ and the soft parameter $\beta_{\phi}$.

The corresponding eigenvectors are given by
\begin{eqnarray}  \label{evpseudoscalars}
G_{1}&=&\frac{1}{\sqrt{1+ \tan^{2} \beta}}\left( - \tan \beta \,\textrm{Im} {H}_{1}^{0}+\textrm{Im} {H}_{2}^{0}\right), \,\
G_{2}= \frac{1}{\sqrt{1+ \tan^{2} \xi}}\left( - \tan \xi \,\textrm{Im} {\phi}_{1}+\textrm{Im} {\phi}_{2}\right),   \nonumber \\
A_{1}&=&\frac{1}{\sqrt{1+ \cot^{2} \beta}}\left( \cot \beta\, \textrm{Im} {H}_{1}^{0}+\textrm{Im} {H}_{2}^{0} \right), \,\
A_{2}=\frac{1}{\sqrt{1+ \cot^{2} \xi}}\left( \cot \xi \,\textrm{Im} {\phi}_{1}+\textrm{Im} {\phi}_{2} \right). \nonumber \\
\end{eqnarray}

\subsection{Neutral scalars}
\label{subsec:neutral}

\begin{figure}[h]
	\centering
	\includegraphics[width=5in]{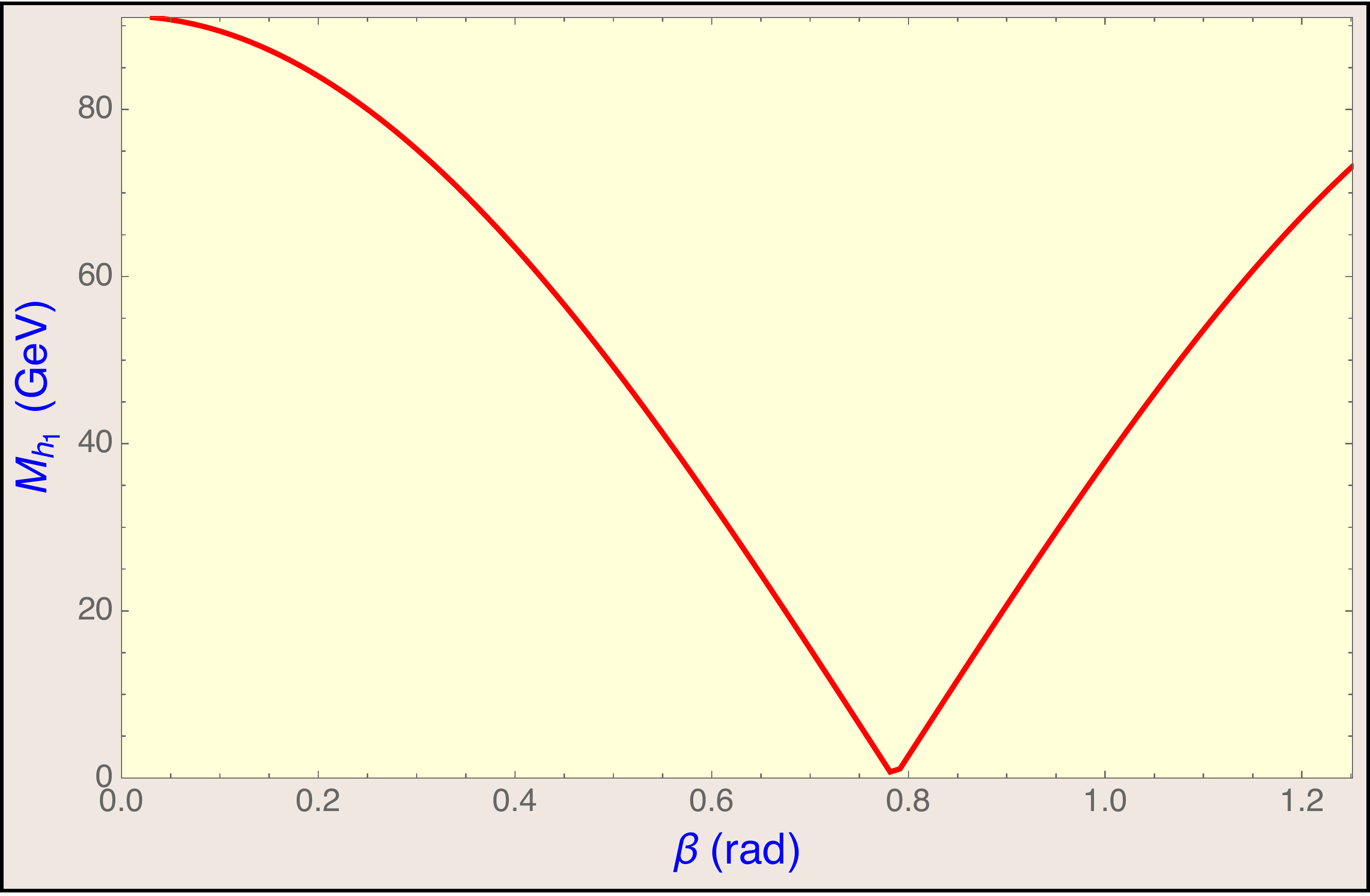}
	\caption{The values of $M_{h^{0}_{1}}$ at the tree level choosing $\beta_{H}= \beta_{\phi}=1$ TeV in terms of $\beta$.}
	\label{fig6}
\end{figure}

The neutral scalar mass matrix in the basis $(\textrm{Re}{H}_{1}^{0},\textrm{Re}{H}_{2}^{0},$ $\textrm{Re}{\phi}_{1},\textrm{Re}{\phi}_{2})$ is given by
\begin{eqnarray}
M^2_S &=& \left(
\begin{array}{cccc}
A & E & - \frac{1}{2}g^{2}_{Y^{\prime}}v_{1}u_{1} & \frac{1}{2}
g^{2}_{Y^{\prime}}v_{1}u_{2} \\
E & B & \frac{1}{2}g^{2}_{Y^{\prime}}v_{2}u_{1} & -
\frac{1}{2}g^{2}_{Y^{\prime}}v_{2}u_{2} \\
- \frac{1}{2}g^{2}_{Y^{\prime}}v_{1}u_{1} & \frac{1}{2}g^{2}_{Y^{\prime}}v_{2}u_{1} & C & F \\
\frac{1}{2}g^{2}_{Y^{\prime}}v_{1}u_{2} & - \frac{1}{2}g^{2}_{Y^{\prime}}v_{2}u_{2} & F & D
\end{array}
\right), \nonumber \\
\end{eqnarray}
where
\begin{eqnarray}
&& A= \frac{1}{4}(g^{2}+g^{2}_{Y^{\prime}})v^{2}_{1}+ \frac{v_{2}}{v_{1}}\beta_{H}^{2}, \quad
B= \frac{1}{4}(g^{2}+g^{2}_{Y^{\prime}})v^{2}_{2}+
\frac{v_{1}}{v_{2}}\beta_{H}^{2}, \nonumber \\ &&
C= (g^{2}_{BL}+g^{2}_{Y^{\prime}})u^{2}_{1}+
\frac{u_{2}}{u_{1}}\beta_{\phi}^{2}, \quad D= (g^{2}_{BL}+g^{2}_{Y^{\prime}})u^{2}_{2}+ \frac{u_{1}}{u_{2}}\beta_{\phi}^{2}, \nonumber \\ &&
E=-\frac{1}{4}(g^{2}+g^{2}_{Y^{\prime}})v_{1}v_{2}-\beta_{H}^{2}, \quad F=-
(g^{2}_{BL}+g^{2}_{Y^{\prime}})u_{1}u_{2}-\beta_{\phi}^{2}.
\end{eqnarray}
The determinant is given by
\begin{eqnarray}
\det \left( M^2_S \right)&\propto&
\left| \sqrt{\tan \beta}-\sqrt{\cot \beta}\right| \cdot
\left| \sqrt{\tan \xi}-\sqrt{\cot \xi}\right|.
\end{eqnarray}
Therefore, we must have $\tan \beta \neq \cot \beta$ and $\tan \xi \neq \cot \xi$ in order to have non vanishing determinant, and then avoiding
one Goldstone boson in this sector. We need to impose that $\tan \beta$ and $\cot \beta$ can not diverge to get
finite masses to the CP even scalars (the same restrictions  hold to the parameter $\xi)$.  Therefore,  we can impose the following constraints
\begin{eqnarray}
0< \beta < \frac{\pi}{2}\,\ \mathrm{rad} \,\ \mathrm{and} \,\ \beta \neq \frac{\pi}{4}\,\ \mathrm{rad},  \quad
0< \xi < \frac{\pi}{2}\,\ \mathrm{rad} \,\ \mathrm{and} \,\ \xi \neq \frac{\pi}{4}\,\ \mathrm{rad},
\end{eqnarray}
Note that in the MSSM there is not this kind of restriction to $\beta$. It can be seen in Fig.~\ref{fig6}, where we show the mass
of the lighest Higgs, in function of the $\beta$ paramenter, at the tree level. In this  model, as in the MSSM, it is possible to show that, at the tree level, the following relation holds
\begin{eqnarray}
M_{h^{0}_{1}}<M_{Z^{0}}\cos 2\beta .
\end{eqnarray}
As in the MSSM,  $M_{h^0_1}$ gets large corrections from top and stop loops, and the Higgs mass becomes
\begin{equation}
M^{2}_{h_1}\equiv M^{2}_{h_1\,\mathrm{tree}}+ \left. \Delta M^{2}_{h_1} \right|_{\mathrm{1-loop}},
\end{equation}
where~\cite{drees,Djouadi:2005gj}
\begin{eqnarray}
\left. \Delta M^{2}_{h_1} \right|_{\mathrm{1-loop}}\approx \frac{3m^{4}_{t}}{4 \pi v^{2}_1}\left[
\log \left( \frac{\tilde{m}^{2}_{t}}{m^{2}_{t}} \right) + \frac{A^{2}_{t}}{\tilde{m}^{2}_{t}} \left( 1-
\frac{1}{12} \frac{A^{2}_{t}}{\tilde{m}^{2}_{t}} \right) \right] \,\ ,
\end{eqnarray}
with $v_1/\sqrt{2}\approx 174$ GeV. We can get the following  upper limit for $h_{1}^{0}$ \cite{Casas:1994us}
\begin{equation}
M_{h^{0}_{1}}<140\, {\mbox{GeV}}.
\end{equation}

\begin{figure}[h]
\centering
\includegraphics[width=5in]{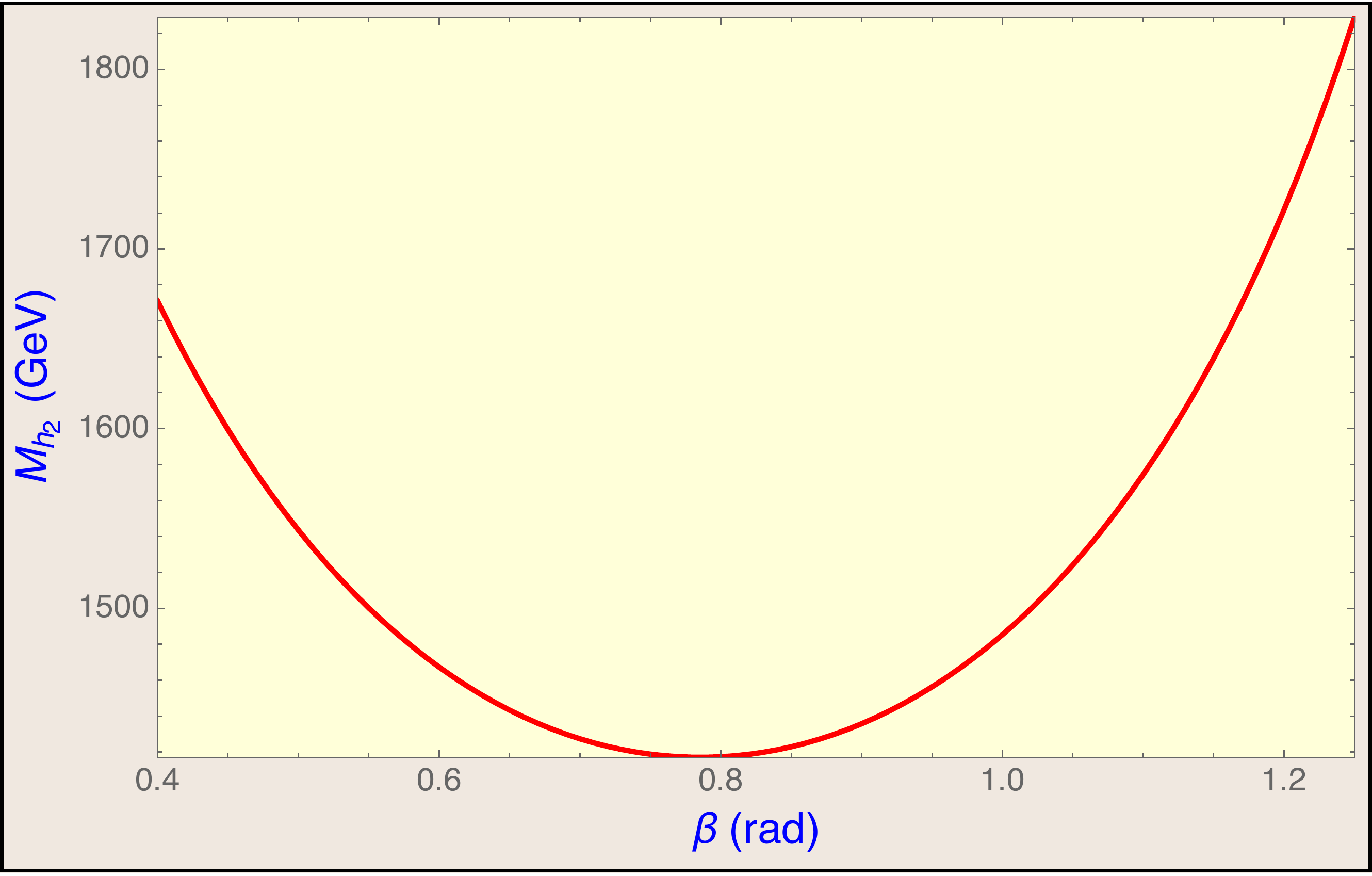}
\caption{The values of $M_{h^{0}_{2}}$ choosing $\beta_{H}= \beta_{\phi}=1$ TeV in terms of $\beta$.}
\label{fig7}
\end{figure}

\begin{figure}[h]
	\centering
	\includegraphics[width=5in]{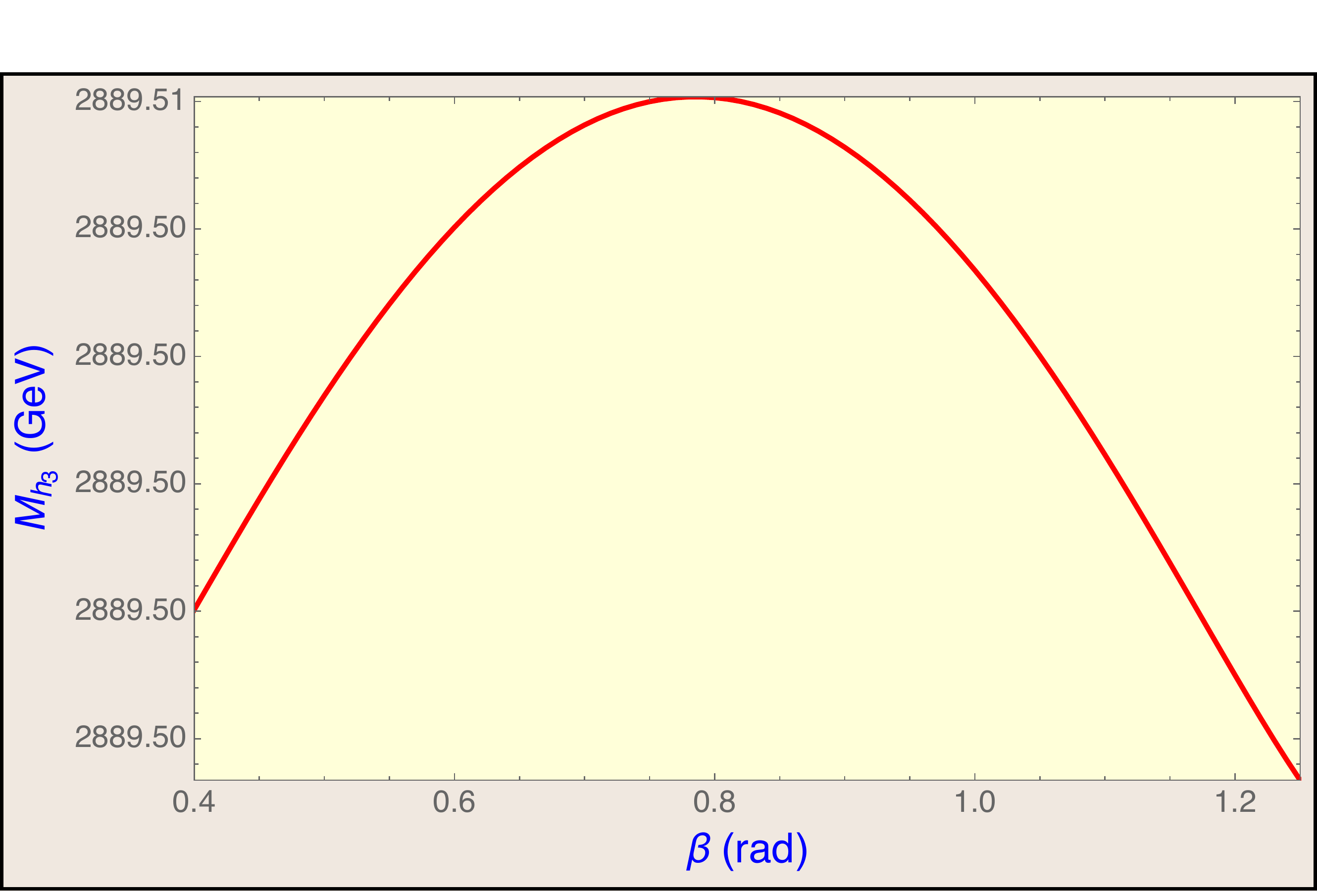}
	\caption{The values of $M_{h^{0}_{3}}$ choosing $\beta_{H}= \beta_{\phi}=1$ TeV in terms of $\beta$.}
	\label{fig8}
\end{figure}

\begin{figure}[h]
	\centering
	\includegraphics[width=5in]{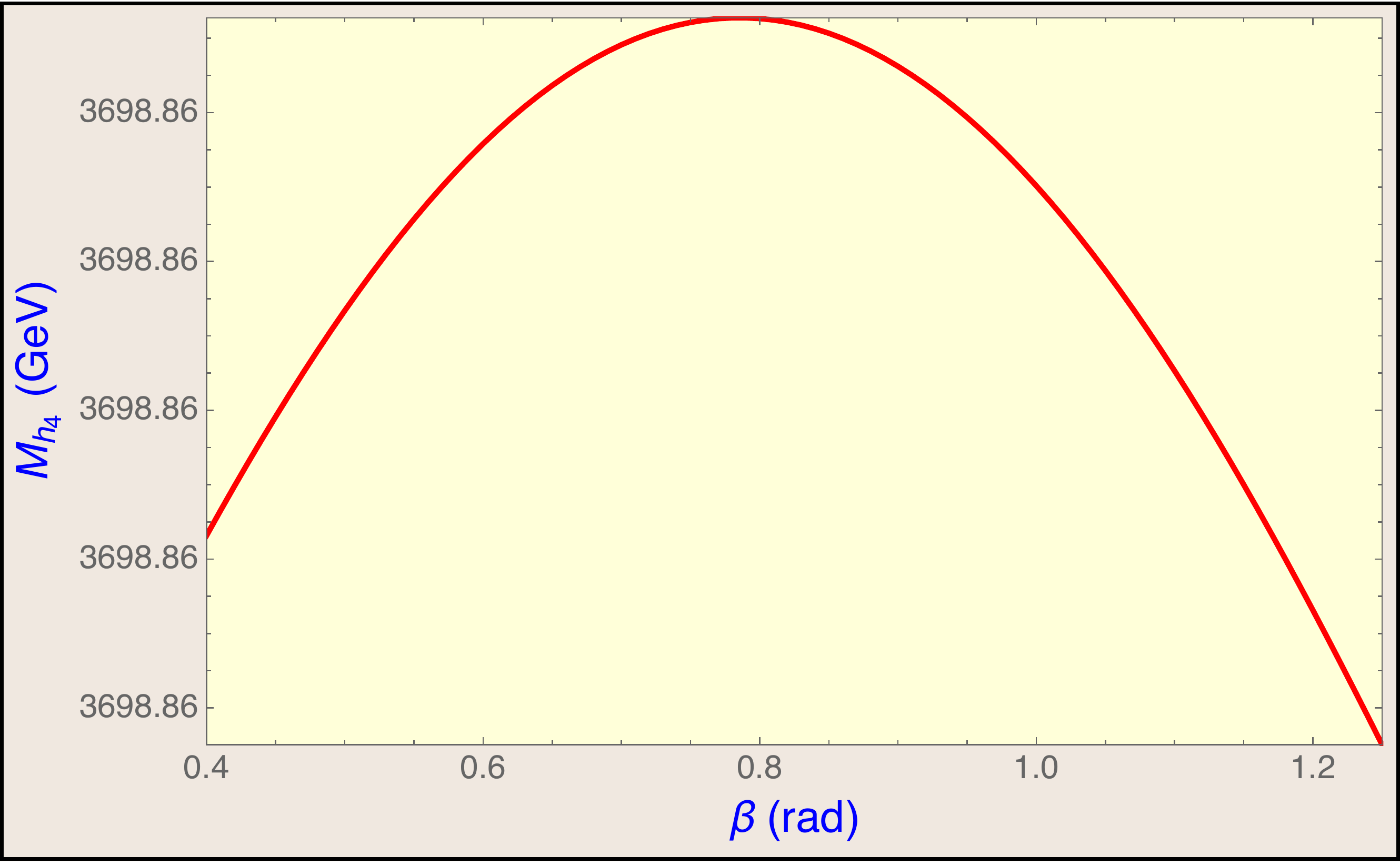}
	\caption{The values of $M_{h^{0}_{4}}$ choosing $\beta_{H}= \beta_{\phi}=1$ TeV in terms of $\beta$.}
	\label{fig9}
\end{figure}

We have also obtained the numerical values to the other scalars. If we choose $\beta =0.4$ rad, and the same parameters used before,
we get the following values
\begin{eqnarray}
M_{h^{0}_{2}}=1620, \,\ M_{h^{0}_{3}}=2889.50, \,\ M_{h^{0}_{4}}=3698.86,
\end{eqnarray}
where all values listed above are in GeV. See Figs.~\ref{fig7}, \ref{fig8} and \ref{fig9}, where these masses are plotted in function of $\beta$ for $\beta_H=\beta_\phi=1\,\textrm{TeV}$.

\section{Flat Directions}
\label{sec:flat}

The universal ingredient in inflation potential is the existence of  a flat direction
which is usually related to a symmetry and/or to a small coupling. The most simple example about flat direction is the Polonyi
model, where only one chiral superfield $\hat{X}(y, \theta)$ is introduced. In this model the K\"ahler potential $K$ and the
superpotential $W$ is given by
$K= \hat{X}^{\dagger}\hat{X}$ and $W= \hat{X}\hat{X}$, respectively. The scalar potential of this model reads
$V= |\lambda|^{2}$, where $\lambda$ is the coupling of quartic term in the scalar potential,
then SUSY is clearly broken for any $\langle X \rangle$ and the latter is a flat direction.
For instance, in the MSSM, $LH_{2}$ is a flat direction if $\langle {\tilde{L}} \rangle=(v_2\,0)^{T}$ and $\langle H_{2} \rangle=( 0\,v_2)^T$~\cite{Gherghetta:1995dv}.

Supersymmetric gauge theories often possess a remarkable vacuum
degeneracy at the classical level. The renormalizable
scalar potential in these models is a sum of squares of $F$-terms and $D$-terms, hence may vanish
identically along certain ``flat directions'' in field space. The properties of the
space of flat directions of a supersymmetric model are crucial for cosmology. As in Ref.~\cite{Allahverdi:2006cx}, this model has also a flat direction given by
\begin{equation}
\phi=\frac{\tilde{L}+H_{1}+\tilde{N}}{\sqrt3}.
\label{flat1}
\end{equation}

The potential at the flat direction is
\begin{equation}
V(\vert\phi\vert)=\frac{m^2_\phi}{2}\vert\phi\vert^2+\frac{f^{\nu}}{12}\vert\phi\vert^4-\frac{Af^{\nu}}{6\sqrt{3}}\vert\phi\vert^3,
\label{potentialflat}
\end{equation}
where the flat direction mass is given by \cite{Allahverdi:2006cx}
\begin{equation}
m^{2}_{\phi}=\frac{m^2_{\tilde{L}}+m^2_{H_1}+m^2_{\tilde{N}}}{3}.
\end{equation}
In this model the inflaton is a gauge-invariant
combination of the right-handed sneutrino, the slepton, and the Higgs field, which generates a flat direction suitable
for inflation if the Yukawa coupling is small enough~\cite{Allahverdi:2007wt,Allahverdi:2006cx}. At the flat direction local minimum $\phi_0=\sqrt{3}\frac{m_{\phi_0}}{f^{\nu}}=X m_\phi\left(\frac{0.05\,\textrm{eV}}{m_\nu}\right)$, the potential is given by

\begin{equation}
V(\vert\phi_0\vert)=\frac{m^4_\phi}{4{f^{\nu}}^2}=Y m^4_\phi \left(\frac{0.05\,\textrm{eV}}{m_\nu}\right)^2,
\label{phi0}
\end{equation}
where $X$ and $Y$ are  proportional to ${1/f^{\nu}}$. In this model $\hat{\phi}_{1}\hat{\phi}_{2}$, $\hat{\phi}_{2}\hat{N}_{i}$ and $\hat{N}_{i}\hat{N}_{j}$ give also flat directions.

The term $\hat{\phi}_{2}\hat{N}_{i}\hat{N}_{j}$ gives also a flat direction in this model and it is similar to
$\hat{L}\hat{H}_{1}\hat{N}$. Using this new flat direction we can easily derive the results presented in
Eq.(\ref{potentialflat}) and Eq.(\ref{phi0}) and, hence,  reproducing the results in Ref.~\cite{Allahverdi:2007wt,Allahverdi:2006cx} without fine tuning.

\section{Conclusions}
\label{sec:con}

We have considered a supersymmetric version of a model in which the $U(1)_Y$ factor in the SM is substituted by $U(1)_{Y^\prime}\otimes U(1)_{B-L}$ which is broken down to $U(1)_Y$ by non-trivial scalar singlets VEVs at an energy scale higher than the electroweak scale~\cite{Montero:2007cd}. We have introduced three right-handed neutrinos, with the same $B-L$ as the active neutrinos, that are singlet with respect to the SM interactions but are not with respect to the $U(1)_{Y^\prime}$ and $U(1)_{B-L}$ ones.
Charged lepton and neutrino masses are generated at the tree level. In the neutrino case, we have implemented a type-I seesaw mechanism. We also have shown the mass spectra of all the other fermions and scalars, and also of the gauge bosons. In particular we have shown that the lightest neutral scalar has a upper limit of $140$ GeV when radiative corrections are taken into account.

The model has interesting features such as i) neutrinos do not mix with neutralinos, ii) the conservation of the $R$-parity is related not to $B$ and $L$  conservation separately, but with the $B-L$ conservation; and iii) the model has several flat directions due to the existence of a gauge-invariant scalar combination of the right-handed sneutrinos, sleptons and one the Higgs doublets.
It has been shown in Ref.~\cite{Allahverdi:2007wt} that a model with a flat direction, as in Eq.~(\ref{flat1}), may unify inflation, dark matter and neutrino masses. However, the price to be paid is a fine tuning in the Yukawa sector. The details of this unification without resorting to a Yukawa fine tuning  will be shown elsewhere~\cite{novo}.

\begin{center}
{\it Acknowledgement}:
\end{center}

MCR would like to thank to IFT-UNESP for its kind hospitality during the preparation of this article. This work was supported (BLSV) by CAPES. VP thanks to CNPq  for partial financial support.

\newpage
\begin{table}[h]
	\begin{center}
		\begin{tabular}{|c|c|c|c|}
			\hline
			$\mbox{ Chiral Superfield} $ &  $\mbox{ Slepton} $ & $\mbox{ Lepton} $ & ${\rm{Auxiliar \,\ Field}}$ \\ \hline
			$\hat{L}_{i}$ & $\tilde{L}_{i}$ & $L_{i}$ &  $F_{L_{i}}$ \\ \hline
			$\hat{E}_{i}$ & $\tilde{E}_{i}$ & $E_{i}$ &  $F_{E_{i}}$ \\ \hline
			$\hat{N}_{i}$ & $\tilde{N}_{i}$ & $N_{i}$ &  $F_{N_{i}}$  \\ \hline
		\end{tabular}
	\end{center}
	\caption{Particle content in each chiral superfield introduced in Eq.~(\ref{leptonsm1}).}
	\label{table1}
\end{table}

\begin{table}[h]
\begin{center}
\begin{tabular}{|c|c|c|c|}
\hline
$\mbox{ Chiral Superfield} $ &  $\mbox{ Squark} $ & $\mbox{ Quark} $ & ${\rm{Auxiliar \,\ Field}}$ \\ \hline
$\hat{Q}_{i}$ & $\tilde{Q}_{i}$ & $Q_{i}$ &  $F_{Q_{i}}$ \\ \hline
$\hat{D}_{i}$ & $\tilde{D}_{i}$ & $D_{i}$ &  $F_{{D}_{i}}$ \\ \hline
$\hat{U}_{i}$ & $\tilde{U}_{i}$ & $U_{i}$ &  $F_{{U}_{i}}$  \\ \hline
\end{tabular}
\end{center}
\caption{Particle content in each chiral superfield given in Eqs.~(\ref{quarkssm1}) and (\ref{quarksm2}).}
\label{table2}
\end{table}

\begin{table}[h]
	\begin{center}
		\begin{tabular}{|c|c|c|c|c|}
			\hline
			$\mbox{ Chiral Superfield} $ &  $\mbox{ Higgs} $ & $\mbox{ Higgsino} $ & ${\rm{Auxiliar \,\ Field}}$ \\ \hline
			$\hat{H}_{1}$ & $H_{1}$ & $\tilde{H}_{1}$ &  $F_{H_{1}}$ \\ \hline
			$\hat{H}_{2}$ & $H_{2}$ & $\tilde{H}_{2}$ &  $F_{H_{2}}$ \\ \hline
			$\hat{\phi}_{1}$ & $\phi_{1}$ & $\tilde{\phi}_{1}$ &  $F_{\phi_{1}}$  \\ \hline
			$\hat{\phi}_{2}$ & $\phi_{2}$ & $\tilde{\phi}_{2}$ &  $F_{\phi_{2}}$  \\ \hline
		\end{tabular}
	\end{center}
	\caption{Particle content in the chiral superfields given in Eqs.~(\ref{scm1}) and (\ref{esc1model}).}
	\label{table3}
\end{table}

\begin{table}[h]
	\begin{center}
		\begin{tabular}{|c|c|c|c|c|c|}
			\hline
			${\rm{Group}}$ & ${\rm Vector \,\ Superfield}$ & ${\rm{Gauge \,\ Boson}}$ & ${\rm{Gaugino}}$ & ${\rm{Auxiliar \,\ Field}}$ & ${\rm Gauge \,\ constant}$ \\
			\hline
			$SU(3)_{C}$ & $\hat{G}^{A}$ & $g^{A}$ & $\tilde{g}^{A}$ & $D_{g}$ & $g_{s}$  \\
			\hline
			$SU(2)_{L}$ & $\hat{W}^{a}$ & $W^{a}$ & $\tilde{W}^{a}$ & $D_{W}$ & $g$ \\
			\hline
			$U(1)_{Y^{\prime}}$ & $\hat{b}_{Y^{\prime}}$ & $b_{Y^{\prime}}$ & $\tilde{b}_{Y^{\prime}}$ & $D_{Y^{\prime}}$ & $g_{Y^{\prime}}$ \\
			\hline
			$U(1)_{B-L}$ & $\hat{b}_{BL}$ & $b_{BL}$ & $\tilde{b}_{BL}$ & $D_{BL}$ & $g_{BL}$ \\
			\hline
		\end{tabular}
	\end{center}
	\caption{Information on fields contents of each vector superfield of this model.}
	\label{table4}
\end{table}

\newpage

\begin{table}[h]
	\begin{center}
		\begin{tabular}{|c|c|c|c|c|c|c|}
			\hline
			${\rm{Fermion}} $ & $L_{i}$ & $E_{i}$ & $N_{i}$ & $Q_{i}$ & $U_{i}$ & $D_{i}$ \\
			\hline
			$B-L$ & $-1$ & $+1$ & $+1$ & $+1/3$ & $-1/3$ & $-1/3$  \\
			\hline
			$R$ & $+1$ & $+1$ & $+1$ & $+1$ & $+1$ & $+1$   \\
			\hline
		\end{tabular}
	\end{center}
	\caption{The $B-L$ quantum number and $R$-parity of each usual fermion in the model.}
	\label{table5}
\end{table}

\begin{table}[h]
	\begin{center}
		\begin{tabular}{|c|c|c|c|c|}
			\hline
			${\rm{Higgsino}}$ & $\tilde{H}_{1}$ & $\tilde{H}_{2}$ & $\tilde{\phi}_{1}$ & $\tilde{\phi}_{2}$ \\
			\hline
			$B-L$ & $0$ & $0$ & $2$ & $-2$   \\
			\hline
			$R$ & $-1$ & $-1$ & $-1$ & $-1$  \\
			\hline
		\end{tabular}
	\end{center}
	\caption{The $B-L$ quantum number and $R$-parity of each higgsino in the model.}
	\label{table6}
\end{table}


\newpage


\end{document}